\documentclass[12pt,a4paper]{article}
\usepackage{color,graphics,amsmath,epsfig,rotating}

\begin{document}

\setlength{\unitlength}{1mm}
\renewcommand{\arraystretch}{1.4}

%------------------------------------------------------------------------
% new definitions, abreviations, etc
%------------------------------------------------------------------------

\def\micromegas      {{\tt micrOMEGAs}}
\def\cpsh            {{\tt CPsuperH}}
\def\micro{\tt micrOMEGAs}

\def\ma{M_A}
\def\ra{\rightarrow}
\def\mneut{m_{\tilde{\chi}^0_1}}
\def\mchi{m_{\tilde{\chi}^0_i}}
\def\mneutt{m_{\tilde{\chi}^0_2}}
\def\mneuth{m_{\tilde{\chi}^0_3}}
\def\mneutf{m_{\tilde{\chi}^0_4}}
\def\mchar{m_{\tilde{\chi}^+_1}}
\def\mchart{m_{\tilde{\chi}^+_2}}
\def\msel{m_{\tilde{e}_L}}
\def\mser{m_{\tilde{e}_R}}
\def\mslo{m_{\tilde{\tau}_1}}
\def\mslt{m_{\tilde{\tau}_2}}
\def\msul{m_{\tilde{u}_L}}
\def\msur{m_{\tilde{u}_R}}
\def\msdl{m_{\tilde{d}_L}}
\def\msdr{m_{\tilde{d}_R}}
\def\msto{m_{\tilde{t}_1}}
\def\mstt{m_{\tilde{t}_2}}
\def\msbo{m_{\tilde{b}_1}}
\def\msbt{m_{\tilde{b}_2}}
\def\sw{s_W}
\def\cw{c_W}
\def\ca{\cos\alpha}
\def\cb{\cos\beta}
\def\sa{\sin\alpha}
\def\sb{\sin\beta}
\def\tb{\tan\beta}
\def\ssi{\sigma^{SI}_{\chi N}}
\def\si{\sigma^{SI}}
\def\sip{\sigma^{SI}_{\chi p}}
\def\ssd{\sigma^{SD}_{\chi N}}
\def\sd{\sigma^{SD}}
\def\sdp{\sigma^{SD}_{\chi p}}
\def\sdn{\sigma^{SD}_{\chi n}}
\def\msl{M_{\tilde l}}
\def\msq{M_{\tilde q}}
\def\bsg{B(b\rightarrow s\gamma)}
\def\bsmu{B(B_s\rightarrow\mu^+\mu^-)}
\def\btau{R(B\rightarrow\tau\nu)}
\def\Omg{\Omega h^2}
\def\sip{\sigma^{SI}_{\chi p}}
\def\amu{\delta a_\mu}
\def\lsp{\tilde\chi^0_1}
\def\neuto{\tilde\chi^0_1}
\def\neuti{\tilde\chi^0_i}
\def\neutt{\tilde\chi^0_2}
\def\neuth{\tilde\chi^0_3}
\def\neutf{\tilde\chi^0_4}
\def\chargi{\tilde\chi^+_i}
\def\charg{\tilde\chi^+_1}
\def\chargt{\tilde\chi^+_2}
\def\gluino{\tilde{g}}
\def\ul{\tilde{u}_L}
\def\ur{\tilde{u}_R}
\def\stau{\tilde{\tau}}
\def\sl{\tilde{l}}
\def\sq{\tilde{q}}

% Def. fuer groesser-ungefaehr:
\newcommand{\gsim}{\;\raisebox{-0.9ex}           {$\textstyle\stackrel{\textstyle >}{\sim}$}\;}

%=======================================================================
% Title
%=======================================================================

\begin{flushright}
   \vspace*{-18mm}
   Date: \today
\end{flushright}
\vspace*{2mm}

\begin{center}

{\Large\bf Constraining the MSSM with universal gaugino masses and implication for searches at the LHC} \\[8mm]

{\large   G.~B\'elanger$^1$, F.~Boudjema$^1$,  A.~Pukhov$^2$, R.~K.~Singh$^{1,3}$}\\[4mm]

{\it 1) LAPTH, Univ. de Savoie, CNRS, B.P.110,  F-74941 Annecy-le-Vieux, France\\
     2) Skobeltsyn Inst. of Nuclear Physics, Moscow State Univ., Moscow 119992, Russia \\
   3) Institut f\"ur Theoretische Physik und Astrophysik, Universit\"at W\"urzburg,\\
D-97074  W\"urzburg, GERMANY
}\\[4mm]

\end{center}

\begin{abstract}
Using a Markov chain Monte Carlo  approach, we find the allowed parameter space of a MSSM model 
with seven free parameters. In this model universality conditions at the GUT scale are imposed on the gaugino sector. 
We require in particular that the
relic density of dark matter saturates the value extracted from cosmological measurements assuming a
standard cosmological scenario. We characterize the parameter space of 
the  model that satisfies experimental constraints and illustrate the complementarity of the LHC 
searches,  B-physics observables and direct 
dark matter searches for further probing the parameter space of the model. 
We also explore the different decay chains expected for the coloured particles that 
would be produced at LHC. 
\end{abstract}

\section{Introduction}

Among the large number of theoretical models proposed to either solve the hierarchy problem and/or explain dark matter
 with a new stable particle, the minimal supersymmetric model (MSSM)   remains one of the favourite. 
 Supersymmetry not only provides a solution to both these problems but also predicts new physics around the TeV scale.  
The main drawback of the MSSM apart from the lack of evidence for supersymmetric particles is the 
large number of unknown parameters most of which describe the symmetry breaking sector. 
With the improved sensitivities of dark matter searches in astroparticle
 experiments~\cite{Angle:2007uj,Ahmed:2008eu,Adriani:2008zr,Abdo:2009zk,Aharonian:2008aaa,Adriani:2008zq}, the precise determination of the DM relic density 
 from cosmology~\cite{Dunkley:2008ie,Spergel:2006hy,Tegmark:2006az}, the latest results from the Tevatron~\cite{Aaltonen:2008my,Abazov:2009zi}
%~\cite{Abazov:2009zi,Glatzer:2008ur} 
and the precision measurements,  large regions of the parameter space of the supersymmetric models are being probed. 
 This will continue  in the near future with a number of direct and indirect detection experiments improving 
 their sensitivities~\cite{Aprile:2009yh,Bruch:2007zz, Moiseev:2008zz,Mocchiutti:2009sj} and most importantly with the LHC starting to take data.  
 The LHC running at the full design energy of 14TeV  offers good 
 prospects for producing coloured supersymmetric particles lighter than 2-3~TeV,
 for discovering one or more Higgs scalars~\cite{atlas_tdr,CMS_tdr}  and for measuring the rare processes in the flavour sector, 
 in particular in B-physics~\cite{Buchalla:2008jp}. Furthermore some properties of the sparticles, in particular mass differences can be measured
 precisely in some scenarios~\cite{atlas_tdr,CMS_tdr}. 
 
The first studies that extracted constraints on supersymmetric models worked in general within the context of the MSSM embedded in a GUT
scale model such as the CMSSM~\cite{Ellis:2007fu,Baer:2003yh,Belanger:2004ag}. After specifying the fundamental model parameters at the high scale, 
the renormalisation group equations are  used to obtain the weak scale particle spectrum.
This approach provides a convenient framework  for phenomenological analyses as the number of free parameters is reduced 
drastically  compared to the general MSSM (from {\cal O}(100) to $4$ and  $1/2$ parameters in the case of the CMSSM). 
The drawback is that one is often confined to very specific scenarios, for example in the CMSSM the LSP is dominantly bino
over most of the parameter space. This has important consequences for the dark matter relic abundance.   
Furthermore it was customary to choose some specific values for some of the  MSSM or even the SM parameters for 
a convenient representation of the parameter space in two-dimensions. While the link between specific observables and allowed region of
parameter space is easier to grasp in this framework, the allowed parameter space appeared much more restrictive than if
all free parameters were allowed to vary.

%~\cite{Profumo:2004at} bssm fit + correlation between observables and susy spectra

In the last few years  efficient methods for exploring multi-dimensional parameter space have been used in particle physics and more
specifically for determining the allowed parameter space of the CMSSM. 
This approach showed that the often narrow strips in parameter space obtained when varying only two parameters at a time 
fattened to large areas ~\cite{Baltz:2004aw,Allanach:2005kz,Allanach:2006cc,deAustri:2006pe} after letting all parameters 
of the CMSSM and the SM vary in the full range. 
With this efficient parameter space sampling method it  becomes possible to relax some 
theoretical assumptions and consider 
the full parameter space of the MSSM. Because the number of  experimental constraints on TeV scale physics is still rather limited 
it seems a bit premature to go to the full fledge ${\cal O}(100)$ parameters of the MSSM or even to the 
19 parameters that characterize the 
model when assuming no flavour structure and equality of all soft parameters for the first and 
second generations of sfermions (for an
approach along these lines see ~\cite{Berger:2008cq,AbdusSalam:2009qd}). 
%RKS: Could not understand the following sentence %gb
Furthermore many parameters, for example those of the first and second
generations of squarks, once chosen  to be equal
to avoid strong flavour-changing neutral current constraints, do not  play an important role in 
the observables  selected to fit the model. Here we  consider a model where input parameters of
 the MSSM are defined
 at the  weak scale  and we add 
some simplifying assumptions: common slepton masses ($M_{\tilde{l}_R}=M_{\tilde{l}_L}=M_{\tilde l}$) 
and common
squark masses ($M_{\tilde{q}_R}=M_{\tilde{q}_L}=M_{\tilde q}$   at the weak scale for all three generations  and universality of 
gaugino parameters at 
the GUT scale. This implies the following relation between the  gaugino masses at the weak scale,
$M_2=2M_1=M_3/3$. We furthermore assume that $A_t$ is the  only non-zero trilinear coupling.
While, as we just argued, the first assumption should not impact much our analysis, the second should certainly be 
considered as a theoretical bias.
This assumption is however well motivated in the context of models defined at the GUT scale.
Most importantly in our approach we keep the higgsino parameter $\mu$ and the gaugino mass $M_2$ 
 as completely independent parameter.
 The relation between the gaugino and higgsino  parameters is what 
determines the nature of the LSP and plays an important role in determining the LSP-LSP annihilation in the early universe. In that sense 
our model 
has many similarities with the non-universal Higgs model which also has $\mu$ and $M_2$ as independent parameters~\cite{Baer:2005bu}.   
%{\bf comment on trilinear}

The observables selected to constrain the model include  the relic density of dark matter, $\Omega h^2$, direct searches for Higgs 
and new particles at colliders,   searches for rare processes such as the muon anomalous magnetic moment as well as various B-physics
observables. Note that the dark matter relic abundance is computed within the standard cosmological scenario. The direct detection of
 dark matter 
while providing stringent constraint on the model introduces additional unknown parameters
both from astrophysics and from strong interactions.  We therefore prefer to consider the direct detection rate as an observable to be 
predicted rather than as a constraint keeping in mind that folding in the astrophysical and hadronic uncertainty 
could however easily introduce an order of magnitude uncertainty in that prediction~\cite{Bottino:2001dj,Belanger:2008sj}.

We find  that each individual parameter of the MSSM model is only weakly constrained, in particular the parameters of the sfermion sector.
The very large allowed parameter space  only reflects the still poor sampling of the total parameter space by experiments.
 The  neutralino sector is better constrained  with a preferred value for the LSP of a few hundred GeV's and a small likelihood
for masses above 900GeV, similarly charginos above 1.2TeV are disfavoured.   We also find  a lower limit on
the pseudo-scalar mass as well as on $\tan\beta$. Furthermore  some correlations between parameters of the model are observed, most notably the
one between $\mu$ and the gaugino mass. This is because those two parameters determine the higgsino content of the LSP.  

After having determined the allowed parameter space, we examined the predictions for direct detection as well as for LHC searches both in 
the Higgs and SUSY sector as well as for B-physics
observables. Although each type of search can only probe a fraction of the total parameter space we find a good 
complementarity between the different searches with less than 10\% of scenarios leading to no signal. For example large
signals for direct detection are expected in the mixed bino/Higgsino LSP scenario that are hard to probe at the LHC.
The LHC searches in the SUSY and Higgs sector are also complementary and B-observables are specially useful in scenarios with  large
$\tan\beta$ and a  pseudoscalar that is not too heavy.  
The predictions for SUSY searches can 
be different from that expected in the constrained CMSSM with in  particular a large fraction of models that only have a 
gluino accessible at LHC,  the squarks being too heavy. To  ascertain how experiments that will take place in the near future 
could further constrain the parameter space of the model we consider specific case studies.   
For example we consider the impact of 
 a signal in $\bsmu$ at Tevatron or of the  observation of a signal in direct detection experiments.
Finally we examine in more details the SUSY signals at the LHC, analysing the preferred decay chains for models that have either a 
gluino or a squark within the reach of the LHC. 
In this analysis we did not include the constraints from indirect detection experiments because the rates predicted feature a strong dependence on
additional quantities such as the dark matter profile or the boost factor. The predictions for the rates for $\bar{p},e^+,\gamma$ will be presented 
 in a separate publication~\cite{belanger_id}.

The paper is organised as follows. The model and the impact of various constraints are described in section 2.
The method used for the fit is described in section 3.  The results of the global fits are presented in section 4 together with  
the impact of a selected number of future measurements. The SUSY signatures are detailed in  section 5. 
The conclusion contains a summary of our results.

\section{Model and constraints}

We consider the MSSM with input parameters defined at the weak scale. We assume minimal flavour violation, equality of the 
soft masses between sfermion generations and 
unification of the gaugino mass at the GUT scale. The latter  leads to $M_2=2M_1=M_3/3$ at the weak scale 
(relaxing this assumption is kept for a further study).
We allow for only one non-zero trilinear coupling $A_t$. For the b-squark the mixing which is $\propto A_b-\mu\tan\beta$
is driven in general by $\mu\tan\beta$ rather than by the trilinear coupling, this approximation is however
not very good in the small sample of   models with $\mu\approx 100$~GeV.   Note also that the Higgs mass at high $\tan\beta$ 
can show some dependence on the sbottom mixing.
For first and second generations of squarks the mixing which depends on fermions masses is negligible except for the  neutralino-nucleon cross 
section since the dominant contributions to the scalar cross section are also dependent on  fermion masses. 
 However since the squark exchange diagram is usually subdominant as compared to Higgs exchange, the neglected contribution of 
  the trilinear coupling falls within the theoretical uncertainties introduced by the hadronic matrix elements~\cite{Belanger:2008sj}. 
Similarly neglecting the the muon trilinear mixing, $A_\mu$, could affect the prediction for $\amu$ but this effect is not large
compared with the  uncertainties on the value extracted from measurements. 
The top quark mass  $m_t$
%,m_b??,\alpha_S??$ {\bf or did we vary only mt?} 
is also used as an input although it has a much weaker influence on the results than in the case of GUT scale models. For the latter 
the top quark mass enters the  renormalization group evolution and can have a large impact on the supersymmetric spectrum in some regions of the parameter space. In the general MSSM  the top quark mass mainly influences the light Higgs mass. We fix $\alpha_s(M_Z)=0.1172$
and $m_b=4.23$~GeV. 
The free parameters of our MSSM model with unified gaugino masses, MSSM-UG, are 
\begin{equation}
\mu, M_2,\msl,\msq,A_t,\tb,M_A,m_t
\label{eq:param}
\end{equation}
The range examined for each of these parameters is listed in Table~\ref{tab:param}.
MSSM-UG has a far more restricted set of paramters than the general MSSM, still this model will show  how the possibilities for SUSY scenarios 
open up. 
%We first perform a likelihood fit of the MSSM at the weak scale to the observables listed in Table
The observables that will be used in the fit are listed in Table~\ref{tab:constraints}. We first review the expectations 
for the role of each observable in constraining the MSSM parameter space.

\begin{table}[!ht]
\caption{\label{tab:param} Range of the free MSSM-UG parameters.} 
\begin{center}
\begin{tabular}{|c|l|l|}\hline
Symbol &  stands for & General range\\ \hline
$\mu$ &  $\mu$ parameter &  $[-3000, 3000]$ GeV\\ \hline
$M_2$ &  Gaugino mass, $2 M_1 = M_2 = M_3/3$ & $[30, 2000]$ GeV\\\hline
$M_{\tilde{l}}$ & Common slepton mass, $M_{\tilde l} = m_{\tilde\ell_{L,R}}$ & $[50, 4000]$ GeV 
\\\hline 
$M_{\tilde{q}}$ & Common squark mass, $M_{\tilde q} = m_{\tilde q_{L,R}}$ & $[50, 4000]$ GeV 
\\\hline 
$A_t$ & Trilinear coupling of $\tilde{t}$ & $[-3000, 3000]$ GeV\\ \hline
$\tan\beta$ & $\tan\beta$ & $[5, 65]$ \\ \hline
$M_A$ & Mass of CP-odd Higgs boson & $[100, 2000]$ GeV\\ \hline
$m_t$ & mass of $t$-quark & $[165,180]$~GeV\\ \hline
\end{tabular}
\end{center}
\end{table}

The most powerful constraint is $\Omega h^2$. Since in the MSSM with gaugino universality there are four different mechanisms 
for efficient neutralino annihilation each calling for a different combination of SUSY parameters, each individual parameter 
is weakly constrained when exploring the full parameter space of the model. The main mechanisms for neutralino annihilation are: 
\begin{itemize}
\item{}
annihilation of a bino LSP into fermion pairs, this necessitates light right-handed sfermions.
\item{} annihilation of a mixed bino/higgsino into W pairs, this means $\mu\approx M_1$. Some contribution from 
chargino and neutralino coannihilations is also expected.
\item{}
annihilation near a (light or heavy) Higgs resonance, a LSP with a non-zero higgsino component is required but the higgsino fraction 
can be very small if $\mneut\approx m_h/2$. For heavy Higgs exchange this process is more efficient at large $\tan\beta$ due to the enhanced couplings of the heavy Higgs to b-quarks and  $\tau$ leptons. 
\item{} Coannihilation with sfermions, this usually means those of the third generation since they are lighter due to mixing. 
Recall that we are
assuming only one slepton mass  $M_{\tilde l}$ and one squark mass $M_{\tilde q}$. 
\end{itemize}

The lower limit on the Higgs mass from LEP strongly constrains the small $\tan\beta$ region.  The upper range for the top quark mass and/or  a large mixing in the stop sector are favoured. The latter means large values of $|A_t|$. 

The muon anomalous magnetic moment still indicates a $3\sigma$ deviation from the SM expectations~\cite{Zhang:2008pka}.
The SUSY  loop contributions can lead to  $\amu>0$ when sparticles are light.  This is the case for
the chargino/sneutrino loop and the neutralino/smuon loop when  $\mu>0$. The latter is enhanced at large values of $\tan\beta$, in this  case it is easier  to accomodate 
$\amu>0$ even with a SUSY spectrum at TeV scale. Recall that we have neglected $A_\mu$, although the mixing is usually dominated by $\mu\tan\beta$,
this introduces some restriction on the parameter space.  

The branching ratio for
$\bsg$ depends on the squark and gaugino/Higgsino sector as well as on the charged Higgs. 
A light pseudoscalar mass is permitted only if squarks are light as well. The squarks then partially cancel the large pseudoscalar contribution but only for $\mu>0$.
Furthermore when $\mu>0$, the chargino contribution  is negative relative to the SM  one
so $\bsg$ can drop too low for light charginos. This is particularly true when $A_t$ is  large and negative, 
meaning a large mixing in the stop sector.

The process $\bsmu$ receives SUSY loops contributions from Higgs exchange as well as from chargino and sfermions. In particular the amplitude for Higgs exchange is enhanced as $\tan^3\beta$. The largest contribution are therefore expected for large $\tb$ and light $\ma$, particularly for light charginos and sleptons. 

In the MSSM, the branching ratio for $B_u\rightarrow \tau\nu_\tau$ is suppressed relative to the standard model prediction by the charged Higgs 
contribution. This contribution is  enhanced as $\tan^2\beta$.  We will use the observable $R(B_u\rightarrow \tau\nu_\tau)$ which gives the ratio between the
SUSY and SM predictions.  
Due to large uncertainties in hadronic matrix elements 
the standard model branching fraction is not known with a good precision, so this observable is not as powerful as other B-physics observables to constrain the parameter space of the MSSM.

LEP limits on the light Higgs and  on sparticles constrain the chargino and neutralino sector as well as the charged sfermions. 
We have not included the recent Tevatron limits on the squark and gluino mass 
 as well as the limits on the chargino mass 
from the trilepton search.  
When presenting our results we will comment on the impact of the new Tevatron results.
\footnote{Note that the trilepton search in $pp\ra \charg\neutt$ extends the LEP constraint on the chargino mass only when sfermions are light, 
no constraint is found in the large scalar mass limit (large $m_0$)   at least in the context of the CMSSM. 
The large fraction of our models which have heavy squarks will therefore not be probed by the Tevatron even if they feature light neutralinos and charginos.}

\section{Parameter sampling method }
In this section we attempt to motivate the MCMC method in a heuristic way,  for
a detailed treatment see Ref.~\cite{mcmc_book}

The likelihood $L$ is the probability distribution function (PDF) $p(d|m)$ for 
a data set $d$ being reproduced by an assumed model $m$. In our case, we assume
the model $m=$MSSM-UG and the data set to be reproduced is given in 
Table~\ref{tab:constraints}. This is a {\em top-down} approach where by varying
the model parameters we find a suitable $m$ to maximise $p(d|m)$. However, in
a {\em bottom-up} approach one would like to know the probability of model $m$ 
being correct once given the data $d$, {\it i.e.}  $p(m|d)$. From
Bayes' theorem we have
%%%%%%%%%%%%%%%%%%%%%%%%%%%%%%%%%
\begin{figure}[!t]
%\begin{minipage}{7.0cm}
\centerline{\epsfig{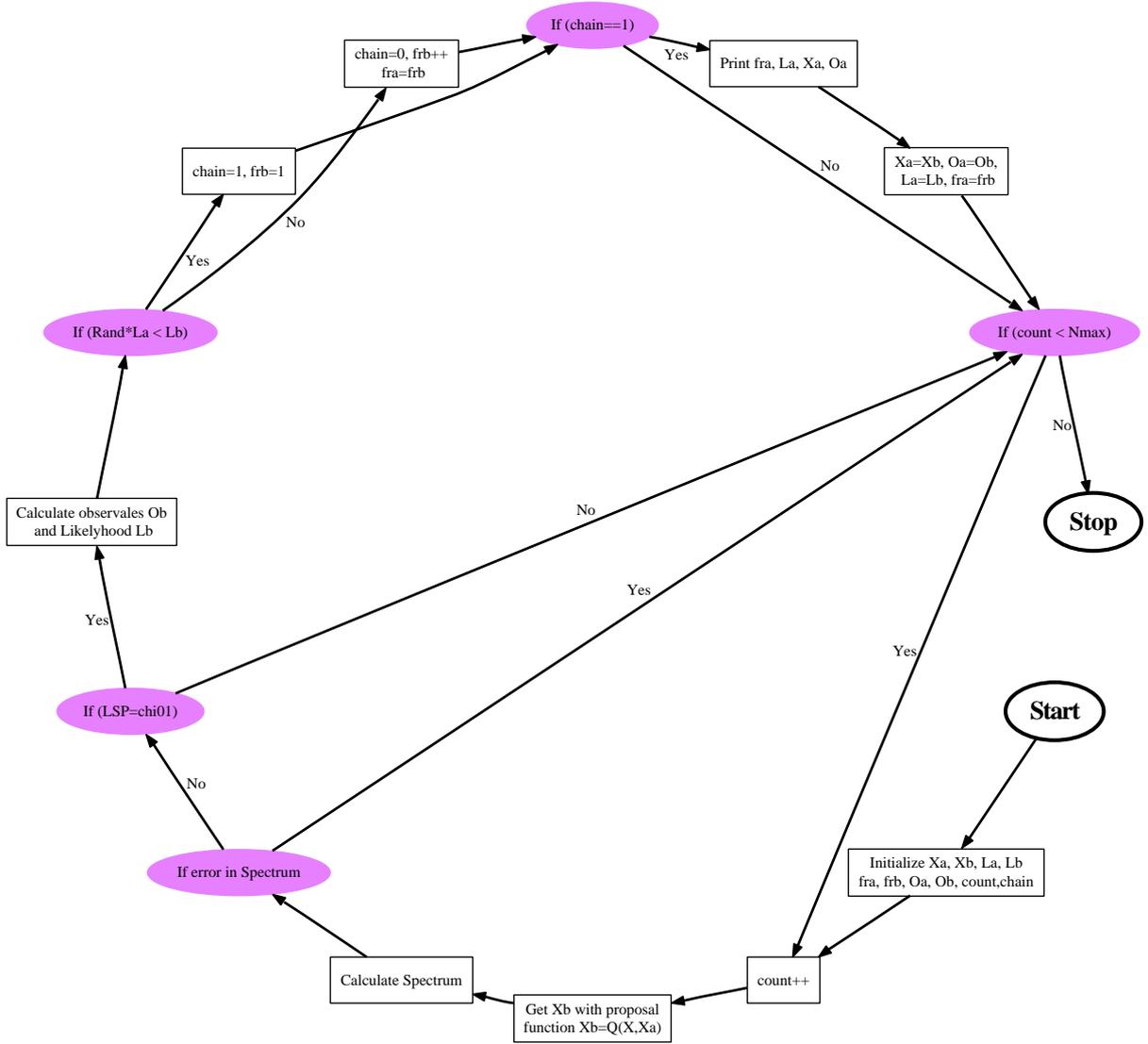}}
%\end{minipage}\hspace{0.5cm}
%\begin{minipage}{8.5cm}
%$Xa,Xb$: The points in the parameter space.\\
%$Oa,Ob$: Observables at points Xa and Xb.\\
%$La,Lb$: Likelihood at Xa and Xb.\\
%fra,frb: Frequency at Xa and Xb.\\[0.5cm]
%The proposal function $Q(X,Xa)$ is a Gaussian centered around $Xa$ for each 
%direction in the parameter space. The variance for each parameter direction 
%differs but it is kept fixed.\\[0.5cm]
%\end{minipage}
\caption{\label{fig:mcmc} Details of the MCMC algorithm used in this paper.
{\tt Xa} and {\tt Xb} are the points in the parameter space, {\tt Oa} and 
{\tt Ob} are the values of the observables at points {\tt Xa} and {\tt Xb}
respectively, {\tt La} and {\tt Lb} are corresponding likelihoods and 
{\tt fra} and {\tt frb} are the frequencies at two points.}
\end{figure}
%%%%%%%%%%%%%%%%%%%%%%%%%%%%%%%%%
$$ p(m|d) = p(d|m) \frac{p(m)}{p(d)},$$
where $p(m)$ is the (absolute) probability of the model $m$ being correct 
and $p(d)$ is the total probability of reproducing the data $d$ for all 
possible models. The $p(d)$ is 
{\em hard} to estimate since it requires knowledge of all possible models. 
Thus, the absolute value of $p(m|d)$ is hard to estimate. However, it is 
possible to estimate the  {\em relative correctedness} of a model $m_1$ 
against another model $m_2$ given the data $d$ by taking the ratio
$$\frac{p(m_1|d)}{p(m_2|d)} = \frac{p(d|m_1) \ p(m_1)}{p(d|m_2) \ p(m_2)}.$$
For simplicity, we take $p(m_1)=p(m_2)$, that is  the probability of two models 
being correct is same or we chose a flat prior over the model space. In this
paper this translates to using a flat prior over the parameter space of MSSM-UG
as listed in Table~\ref{tab:param}. Thus with flat prior we have
$$\frac{p(m_1|d)}{p(m_2|d)} = \frac{L(m_1)}{L(m_2)}.$$
This clearly hints that if we sample the parameter space using a directed 
random walk with transition probability proportional to 
$\min(1,L(m_1)/L(m_2))$, 
the sampling density will be proportional to the likelihood ratio or the 
relative correctedness of the model. The details can be found
in ~\cite{mcmc_book} and a representation of the algorithm used is given in Fig.~\ref{fig:mcmc}.

There are two points in order: For the sampling density to be proportional to
the ratio of the likelihoods the proposal PDF, $Q(\mathbf{X},\mathbf{X_a})$,
should be symmetric about $\mathbf{X_a}$, where  $\mathbf{X},\mathbf{X_a}$ are
point in the parameter space. We use the Gaussian distribution in each 
parameter direction and the proposal PDF looks like:
$$Q(\mathbf{X},\mathbf{X_a}) = \prod_i \exp\left[\frac{-({X^i}-
{X_a^i})^2}{2 \ (\Delta X^i)^2}\right].
$$
The second point is about the definition of likelihood.
The likelihood $L$ is a product of likelihood for each of the observables. For
the observable with definite measurement the likelhood function is a Gaussian
$$\mathbf{G}(O,O_{exp},\Delta O) = \exp\left[\frac{-(O-O_{exp})^2}{2 \ 
(\Delta O)^2}\right].$$
Here, $O_{exp}$ is the central value of the observable, $\Delta O$ is the 
1$\sigma$ error and $O$ is the value of the observable at the proposed point 
$\mathbf{X}$ in the parameter space.
For observables with upper/lower limits the likelihood function is a smooth 
step-like fuction
$$\mathbf{F}(O,O_{exp},\Delta O) = \frac{1}{1+\exp[\pm(O-O_{exp})/\Delta O]}$$
The $+$ve sign is for the lower limit and the $-$ve is for the upper limit.
Here $O_{exp}$ is the 95\% exclusion limit and $\Delta O$ is about 1\% of 
$O_{exp}$ to roughly emulate the 95\% exclusion limit. The list of observables
used to calculate the likelihood is given in Table~\ref{tab:constraints} 
alongwith the correponding experimental values or limits.

\section{Results}

We have first scanned linearly over the 8 parameters of the MSSM-UG using the
Markov Chain Monte Carlo (MCMC) method just described.
The particle spectrum was computed by SoftSusy2.0~\cite{Allanach:2001kg} and fed to 
micrOMEGAs2.2~\cite{Belanger:2004yn,Belanger:2006is,Belanger:2008sj} 
for the computation of all  DM observables as well as of constraints on the parameters of the supersymmetric model.
The SM value for the branching $\bsg$ calculated in ~\cite{Belanger:2004yn} was shifted in order to match the NNLO result of ~\cite{Misiak:2006zs}. 

The range chosen for all parameters is listed  in Table~\ref{tab:param} and 
the six observables used in the fit  listed in Table~\ref{tab:constraints}.
In addition we have imposed the LEP limits on sparticles ($\mchar$, $m_{\tilde{l}}...$) as defined in micrOMEGAs 
and for the likelihood we have assumed a binary step function. 
We then examined the allowed regions for each  input parameter
as well as the predictions for physical parameters and observables. 
%{\bf if not explained before explain contours and likelihood}
We have not imposed the Tevatron limit on gluino and squarks $m_{\gluino}>308$~GeV and $m_{\tilde
q}>379$~GeV~\cite{Abazov:2007ww}.
However after fitting the model  we have checked a posteriori that these constraints did not affect much our analysis.
 Because of the
universality condition the gluino mass limit is always satisfied while the squark limit
is not satisfied in less than one per-mil of our scenarios.
 
\begin{table}[!ht]
\caption{\label{tab:constraints}Observables used in the fit.} 
\begin{center}
\begin{tabular}{|c|c|l|}\hline
Observable & Limit & Likelihood function \\ \hline
$\amu$ & $(27.8 \pm 8.5) \times 10^{-10}$ & $\mathbf{G}
(x,27.8 \times 10^{-10}, 8.5 \times 10^{-10})$\\ \hline 
$B(b\to s\gamma)$ & $(3.55 \pm 0.24) \times 10^{-4}$ &  $\mathbf{G} 
(x,3.55 \times 10^{-4}, 0.24 \times 10^{-4})$\\ \hline
$\Omega h^2$ & $0.113 \pm 0.0105$& $\mathbf{G}(x,0.113,0.0105)$ \\ \hline
$B(B_s\to \mu^+\mu^-)$ & $\le 0.8 \times 10^{-7}$ & $\mathbf{F}(x,0.8 \times
10^{-7}, -0.8 \times 10^{-9})$ \\ \hline
$R(B\to \tau^+\nu)$ & $1.11\pm 0.52$ & $\mathbf{G}(x,1.11, 0.52)$ \\ \hline
%$\sigma(\chi p \to X)$ & $\le 3.0 \times 10^{-8}$ pb& $\mathbf{F}(x,3.0 \times
%10^{-8}, -3.0 \times 10^{-10})$ \\ \hline
$m_h$ & $\ge 114.5$ GeV & $\mathbf{F}(x,114.5,0.6)$ \\ \hline
$m_t$ &   $171.4\pm 2.1$~GeV & $\mathbf{G}(x,171.4, 2.1)$ \\ 
\hline
\end{tabular}
\end{center}
\end{table}

\begin{figure}[!ht]
\setlength{\unitlength}{1mm} \centerline{\epsfig{file=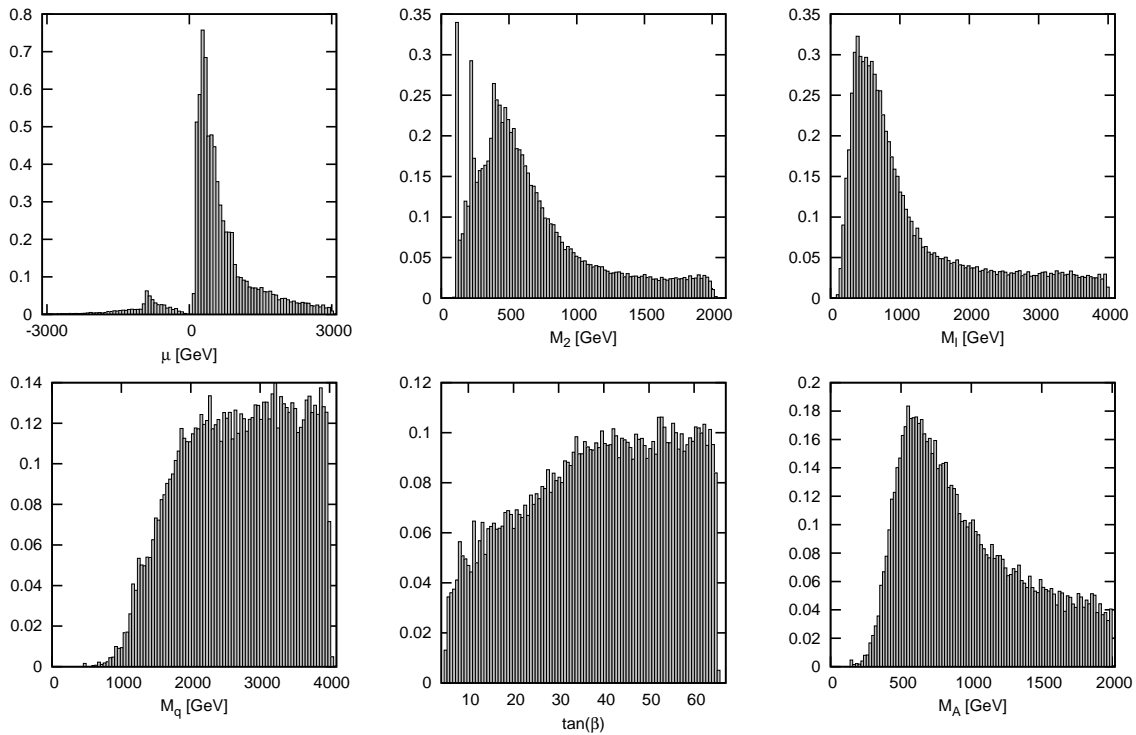,width=10cm,angle=-90}}
\caption{Likelihood function for the model parameters, from top left:
 $\mu$,$M_2$,$\msl$,$\msq,\tb$,$M_A$, all dimensionfull parameters in GeV.} 
\label{fig:1d}
\end{figure}

\begin{table}[!ht]
\caption{Range of the MSSM parameters.} 
\label{tab:cl}
\begin{center}
\begin{tabular}{|c|c|c|c|c|}\hline
Parameter & \multicolumn{2}{|c|}{68\% C.L.} &\multicolumn{2}{|c|}{95\% C.L.}\\\hline
 $\mu$           &$      197.7   $& $     1193    $  &$    -983.5   $& $   2471 $ \\
 $M_2$           &$       259.8  $& $     1077    $  &$     119.6   $& $   1845 $ \\
 $M_{\tilde l}$  &$       398.4  $& $     2270    $  &$     225.4   $& $   3700 $ \\
 $M_{\tilde q}$  &$       1824   $& $     3602    $  &$    1236     $& $   3938 $ \\
 $A_t$           &$      -1735   $& $     2239    $  &$   -2785     $& $   2876 $ \\
 $\tan\beta$    &$        19.5  $& $     65.0    $  &$   8.15      $& $   65.0 $ \\
 $M_A$           &$       537.1  $& $     1489    $  &$     370.2   $& $   1904 $ \\
 $m_t$           &$       169.6  $& $     173.7   $  &$     167.6   $& $   175.6 $ \\
\hline
\end{tabular}
\end{center}
\end{table}

\subsection{The allowed parameter space of the MSSM}

We found wide allowed regions for each model parameter, the distributions are displayed in 
fig.~\ref{fig:1d} and the 68\% and 95\% limits in Table~\ref{tab:cl}.

\begin{itemize}
\item{} As expected, $\mu>0$ is preferred,  this is mainly due to the $\bsg$ and $\amu$ constraints. 
Note however that one can get a reliable global  fit even with  $\amu=0$ so that the
95\%C.L. extends to the negative $\mu$ region, see Table~\ref{tab:param}. 
% The upper limit is a consequence of
%the $\Omega h^2$ constraint which favours a Higgsino component for the LSP especially if the LSP is near the TeV scale. 
\item{} The gaugino mass reaches almost the maximum value probed, $M_2<1845$~GeV at 95\%C.L. 
Charginos above the TeV scale 
are somewhat disfavoured (see  the 68\% C. L. at 1.1~TeV) because they give little contribution to $\amu$, furthermore the light neutralinos LSP  annihilate more efficiently 
except when they have a large Higgsino component or can annihilate near a resonance. 
\item{} The slepton mass distribution is peaked at 500GeV with a long tail that extends to 
almost the upper limit of the region scanned. The sleptons just above the LEP limit are disfavoured with 
$225{\rm GeV}<\msl<3700$~GeV at 95\%C.L.  This shows that although light sleptons
can contribute to the annihilation of a bino LSP and to $\amu$, present data can be 
accomodated with heavy sleptons.  
\item{} Heavy squarks  are preferred with $\msq>1.24$~TeV at 97.5\%C.L. 
 Quarks below the TeV scale tend to give too large corrections to B-physics observables and in particular to  
 $\bsg$. Although the quark contribution can be compensated by that of a light pseudoscalar,  this requires fine-tuning
 and has a small likelihood.
 There is no upper limit on the squark mass as it  can reach  the upper limit of the range used in the scan, 4~TeV.
\item{} Light pseudoscalars are disfavoured with $M_A\geq 370$~GeV at 97.5\% C.L. While it is possible to have a  good fit to 
the data with lighter pseudoscalar masses this occur only after fine tuning  
the sfermion masses in order to satisfy both $\amu$ and $\bsg$. Therefore these models 
have a small likelihood. 
\item{}  The distribution for $\tb$ is skewed towards the upper range and $\tb>8.15$ at 97.5\% C.L. Large values of  $\tb$
 make it easier to escape the LEP bound on the Higgs mass, facilitate annihilation of the LSP through Higgs exchange and can also 
 help explain $\amu$.  
 \item{} There are no preferred value for  $A_t$ although the distribution is skewed towards $A_t>0$,
 see Table~\ref{tab:cl}.
\end{itemize}

A few correlations among input parameters are observed. These are driven to a large extent by the
relic density constraint. First $\mu$ and $\msl$ are correlated: a  large $\mu$ requires 
light sleptons while heavy sleptons need  $\mu$ not much above the TeV scale. 
When sleptons are heavy  the LSP must have a higgsino component to annihilate efficiently. Conversely 
when $\mu$ is large and the LSP is bino, light sleptons are needed for efficient annihilation into 
fermion pairs.  Second $M_2$ is also correlated with $\mu$ and $M_A$:  when  $M_2$ exceeds 1.2TeV,
one needs either  $|\mu|\approx M_2/2\approx M_1$ or $M_2\simeq M_A$. The first correlation is strong
and means that a heavy LSP with a significant higgsino component annihilates efficiently.    
The second correlation  corresponds to $m_{\neuto}\approx M_A/2$
%$M_A$ and $M_2$ are weakly correlated in the sense that the region where 
%$M_2\simeq M_A$ extends to large values of $M_2$. This in fact corresponds to  $m_{\neuto}\approx M_A/2$
with the LSP annihilating near a heavy Higgs resonance. This correlation is weak as we have just mentionned,  other 
large values of $M_2$ are allowed when the LSP has a large higgsino component.

\begin{figure}[!t]
\setlength{\unitlength}{1mm} \centerline{\epsfig{file=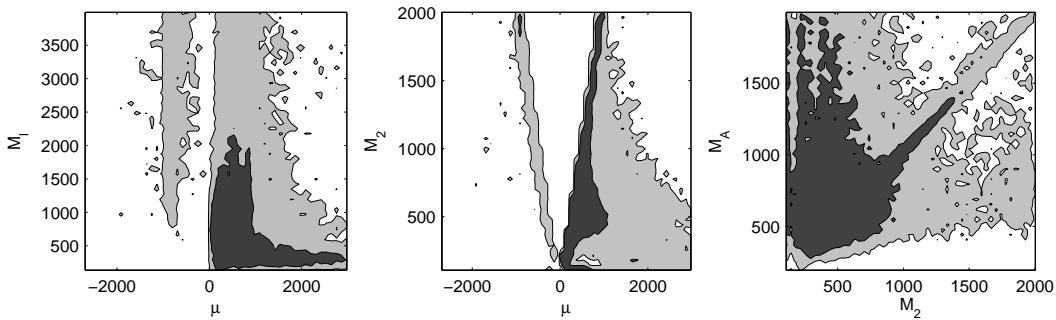,width=14cm}}
\caption{2d-likelihood functions for a) $\msl-\mu$ b)$M_2-\mu$ c)$M_A-M_2$ showing the 68\%C.L.(dark grey) and
95\% C.L. (light grey). All parameters in GeV.}
\label{fig:2d_param}
\end{figure}
% allowed and the whole parameter 
%space for $\msl$ and $M_2$ is allowed even though the xx C.L. contour is limited to $M_2<1,%\msl<1.5$~TeV.   

\subsection{The  particle spectrum}

From this allowed parameter space we can determine the favoured region for the masses of new particles that can be searched for at LHC.

\begin{table}[!ht]
\label{tab:mass}
\caption{The 68\% and 95\% allowed range for sparticle masses and for the higgsino fraction.} 
\begin{center}
\begin{tabular}{|c|c|c|c|c|}\hline
Mass & \multicolumn{2}{|c|}{68\% C.L.} &\multicolumn{2}{|c|}{95\% C.L.}\\\hline
 $m_h$                               &$       117.6  $& $     122.6 $  &$   115.3    $& $    124.8 $ \\
 $m_H$                               &$       537  $& $     1489  $  &$     370    $& $   1900 $ \\
 $\mneut$                            &$       52.2  $& $     522  $  &$      52.2    $& $    874 $ \\
 $\mneutt$                           &$       104  $& $     756  $  &$     104    $& $   1180 $ \\
 $\mneuth$                           &$       269  $& $     1300  $  &$     158    $& $   2500 $ \\
 $\mneutf$                           &$       412  $& $     1650  $  &$     265    $& $   2510 $ \\
 $\mchar$                            &$       103  $& $     749  $  &$     103    $& $   1180 $ \\
 $\mchart$                           &$       410  $& $     1640  $  &$     262    $& $   2490 $ \\
 $\msel$                             &$       410  $& $     2280  $  &$     235    $& $   3710 $ \\
 $\mser$                             &$       406  $& $     2270  $  &$     236    $& $   3700 $ \\
 $\mslo$                           &$       103  $& $     2260  $  &$     103    $& $   3690 $ \\
 $\mslt$                           &$       457  $& $     2280  $  &$     276    $& $   3710 $ \\
 $\msul$                           &$      1840  $& $     3650  $  &$    1230    $& $   3990 $ \\
 $\msur$                           &$      1850  $& $     3660  $  &$    1240    $& $   4000 $ \\
 $\msdl$                           &$      1850  $& $     3670  $  &$    1240    $& $   4010 $ \\
 $\msdr$                           &$      1850  $& $     3660  $  &$    1240    $& $   4000 $ \\
 $\msbo$                           &$      1820  $& $     3620 $  &$     1200    $& $   3960 $ \\
 $\msbt$                           &$      1850  $& $     3650  $  &$    1250    $& $   3990 $ \\
 $\msto$                            &$      1770  $& $     3590  $  &$    1130    $& $   3930 $ \\
 $\mstt$                            &$      1880  $& $     3660  $  &$    1300    $& $   4000 $ \\
 $m_{\tilde{g}}$                     &$       952  $& $     3320  $  &$     478    $& $   5400 $ \\
 $f_H$                               &$ 1.58 \ 10^{-3}$& $     0.297    $  &$ 3.39 \ 10^{-4}$& $  0.659    $ \\
 \hline
\end{tabular}
\end{center}
\end{table}

\begin{figure}[ht]
\centering
\includegraphics[width=14cm,angle=-90]{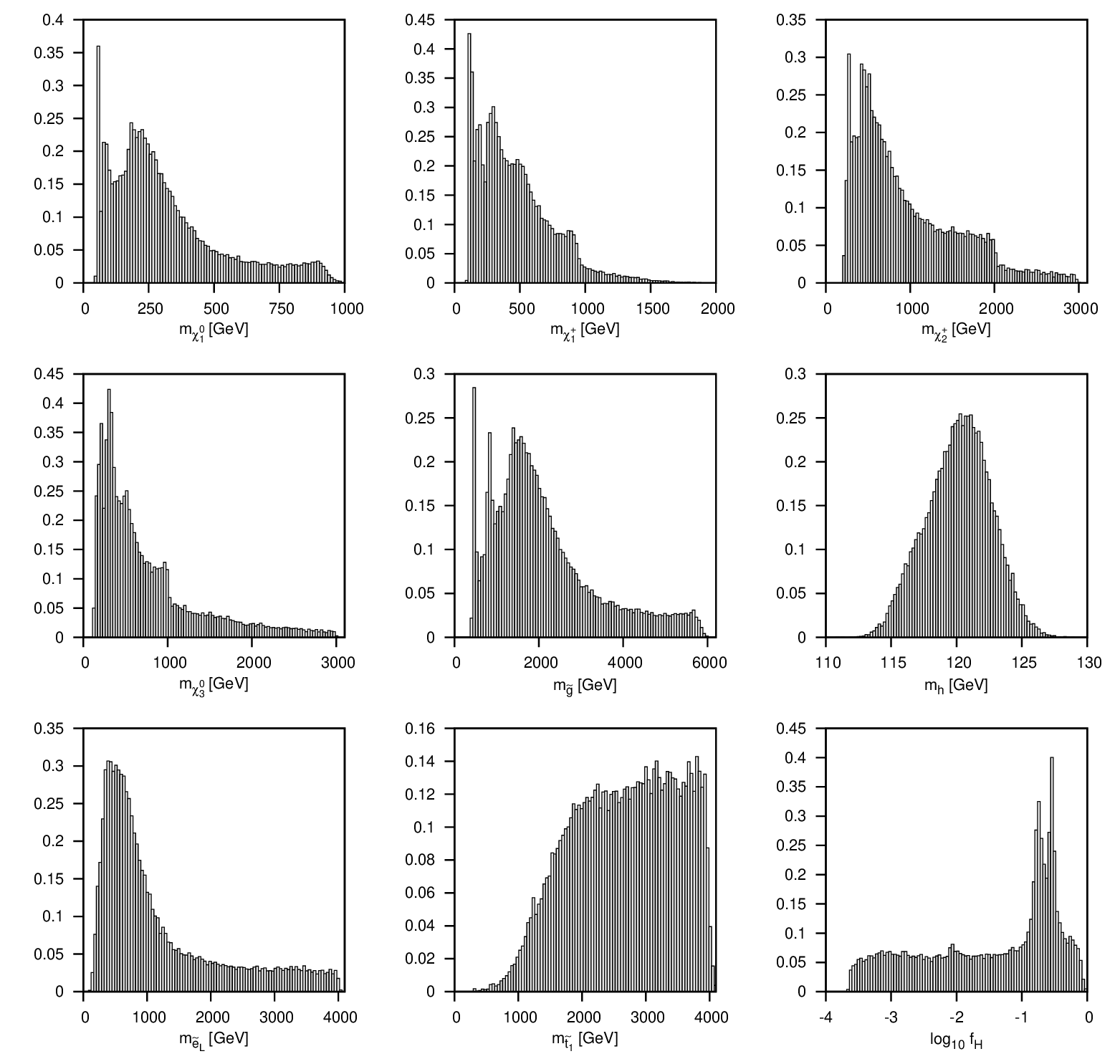}
\caption{Distributions for sparticle masses: $\mneut,\mchar,\mneuth,\mchart$,
$m_{\tilde{g}}$, $m_h,\msel,\msto$ and for the  higgsino fraction, $f_H$. There is a scale factor of  $10^6$ on the y-axis.}
\label{fig:mass}
\end{figure}

\begin{itemize}
\item{} The light Higgs is SM like and $m_h=120.2^{+4.6}_{-4.9}$ GeV at $95\%$C.L.
\item{} The  LSP mass is in the range $52~{\rm GeV}<\mneut<873$~GeV at 95\%C.L.  The lower bound results from the 
assumption of
GUT scale gaugino mass universality and the LEP limit on charginos. The peak in the distribution around 55~GeV corresponds to LSP
annihilation near a light Higgs resonance. 
\item{}The lightest chargino (as well as $\neutt$) could be just above the LEP reach and lies below
$\mchar<1182$~GeV at $97.5\%$C.L. There is a peak in the distribution near 100GeV which is correlated with the LSP peak
correponding to annihilation near the light Higgs.  
\item{} The gluino mass is related to the neutralino mass via the universality condition, the 97.5\%C.L. lower  limit gives
%RKS: A little change is made below.
$m_{\gluino}>477$~GeV and the maximum can exceed 5~TeV. However we have $M_{\tilde g}<3.3$~TeV with 84\%~C.L., which is slightly beyond the reach of the LHC with 
maximum luminosity.
\item{} As mentionned before, squark masses are in the TeV range. Because of the universality assumption on squark masses the mixing in the
stop sector implies that the stop is the lightest squark. We find    $m_{\tilde t_1}>1130$~GeV. Squark 
masses can reach all the way to the upper end of the range of the scan, 4TeV.  There is no guarantee that squarks are within the reach of the LHC.
\item{} Slepton masses reach almost all the way to 4TeV. Since masses of sleptons and gauginos are uncorrelated, 
a large fraction of scenarios have   $m_{\tilde{l}_{1,2}}>m_{\chi_2^0},\mchar$ (around $75\%$ of allowed scenarios). For these scenarios
the preferred decay channel of $\neutt$ is three body and has a branching fraction 
around 3\% into each flavour of charged lepton. In this model all  slepton masses are identical  except for mixing effect so only one mass is
displayed in Fig.~\ref{fig:mass}
\end{itemize}

The higgsino fraction of the LSP strongly influences the properties of the DM, $f_H=|N_{13}|^2+|N_{14}|^2$
where the neutralino mixing matrix is defined in the basis
$\neuto=N_{11} \tilde{B}+N_{12}\tilde{W}+N_{13}\tilde{H_1}+N_{14} \tilde{H_2}$.
The higgsino fraction spans a wide range  $3.4\times 10^{-4}<f_H<0.66$ at 95\%C.L. The distribution is peaked around $0.15<f_H<0.35$.
 A small higgsino fraction is found  either for light LSP's and for LSP's
annihilating near a  Higgs resonance. The higgsino fraction has an impact on the spectrum and in particular on the mass difference 
$\mchar-\mneut$. A large mass difference
is expected for a small higgsino fraction (since in that case $\mchar/\mneut \approx  2$) whereas when $f_H>0.15$, $\mchar-\mneut<M_W$ 
and $\mneutt-\mneut<M_Z$.  This means that for a large number of scenarios that satisfy this condition the decay channels $\charg\ra \neuto W$, 
$\neutt\ra \neuto Z$  are forbidden and only 3-body decays are possible since sleptons are often heavier than the chargino.

\begin{table}[!ht]
\label{tab:obs}
\caption{Range of predictions for observables} 
\begin{center}
\begin{tabular}{|c|c|c|c|c|}\hline
Observable & \multicolumn{2}{|c|}{68\% C.L.} &\multicolumn{2}{|c|}{95\% C.L.}\\\hline
 $\Omega h^2$                          &$     0.102     $& $    0.123     $  &$   0.091       $& $  0.133    $ \\
 $\amu \times 10^{10}$            &$     1.49      $& $    27.2      $  &$  -1.45        $& $  37.5     $ \\
 $B(B\rightarrow X\gamma)\times10^4$   &$     3.19      $& $    3.62      $  &$   3.08        $& $  3.86     $ \\
 $\bsmu\times 10^9$  &$    1.82      $& $   3.31      $  &$  1.48        $& $  20.4     $ \\
 $R(B \rightarrow \tau \nu)$           &$     0.698     $& $   0.981      $  &$   0.382       $& $  0.997    $ \\
 $\sip$ (pb)              &$   2.51\times 10^{-10}      $& $   6.76\times 10^{-8}      $
   &$     9.77\times 10^{-12}    $& $ 2.24 \times 10^{-7}  $ \\
% $log_{10} \sigma^n_{SI}$              &$    -9.58      $& $   -7.16      $  &$     -11.00    $& $ -6.64     $ \\
\hline
\end{tabular}
\end{center}
\end{table}

\subsection{Other observables}

Form the allowed parameter space we then examine the predictions for the observables used in the fit 
together with the direct detection cross-section. 

\begin{figure}[!ht]
\centering
\includegraphics[width=14cm,angle=0]{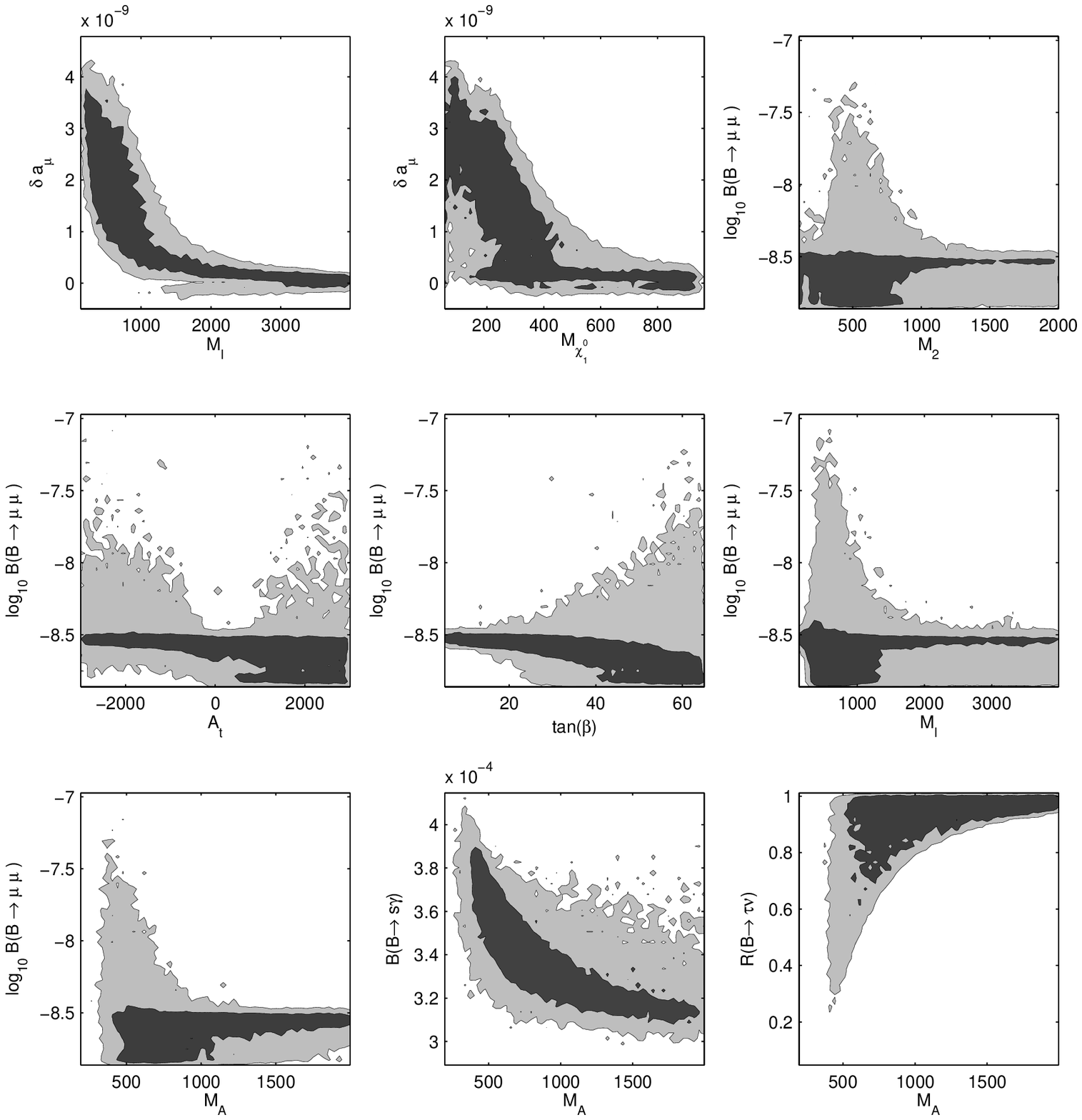}
\caption{2D contours  for some observables, from top left to bottom right:
 a) $\amu-\msl$ b) $\amu-\mneut$  c) $\bsmu-\msl$ d) $\bsmu-A_t$ e)$\bsmu-\tb$ f) $\bsmu-M_A$  g)$\bsmu-M_2$ 
h)$\bsg-M_A$ i) $\btau-M_A$. The 68\%C.L. is in dark grey and 95\%C.L. in light grey.}
\label{fig:2d}
\end{figure}

\begin{itemize}

\item{} The observables $\Omega h^2$ and $\bsg$ are well distributed around the central values used in the fit. 
As expected, $\bsg$ lies near the upper end of the allowed range for small $M_A,\mu, M_2$ and $\msl$ and are
enhanced at very large values of $\tan\beta$. 
\item{} The branching ratio $\bsmu$ is peaked around $2.5\times 10^{-9}$ although a long tail extends to
$2.\times 10^{-8}$ at $95\%$C.L. 
Values near the upper limit are found for $M_A<600$~GeV, 
$\tan\beta>50$,  for not too heavy charginos, $\mchar<750$~GeV and for a LSP with a small higgsino fraction, $f_H<0.05$.
%that is for  $\mu<1$~TeV and $500<M_2<700?$~GeV o

\item{} At 97.5\%C.L., $\btau>0.38$, as for other B-physics observables, the largest suppression to  $\btau$ are found for  small $M_A,\mu, M_2$ and 
$\msl$ as well as for  $\tan\beta>50$.
\item{} The muon anomalous magnetic moment extends over more than an order of magnitude 
and in particular can  include negative values. The peak of the distribution lies near $10^{-10}$, much below the central 
experimental value
we have used indicating that our model does not provide a good fit to this observable. 
Large deviations in $\amu$ are found for light sleptons and neutralinos/charginos, this is why a better fit to the data is found for
low values of $M_2$ and $\msl$. This is illlustrated in Fig.~\ref{fig:2d}a,b. This observable also plays a role in setting the upper limit on the neutralino/chargino masses.
\end{itemize}

%\begin{figure}
%\setlength{\unitlength}{1mm} \centerline{\epsfig{file=fig1/Fig3p_hist.ps,width=14cm}}
%\caption{Sparticles and higgs spectrum : likelihood for selection of neutralino/chargino masses
%gluino?, electronL, uL,t1 + higgsino fraction}
%\label{fig:mass_hist}
%\end{figure}

\begin{figure}[ht]
\centering
\includegraphics[width=14cm,angle=-90]{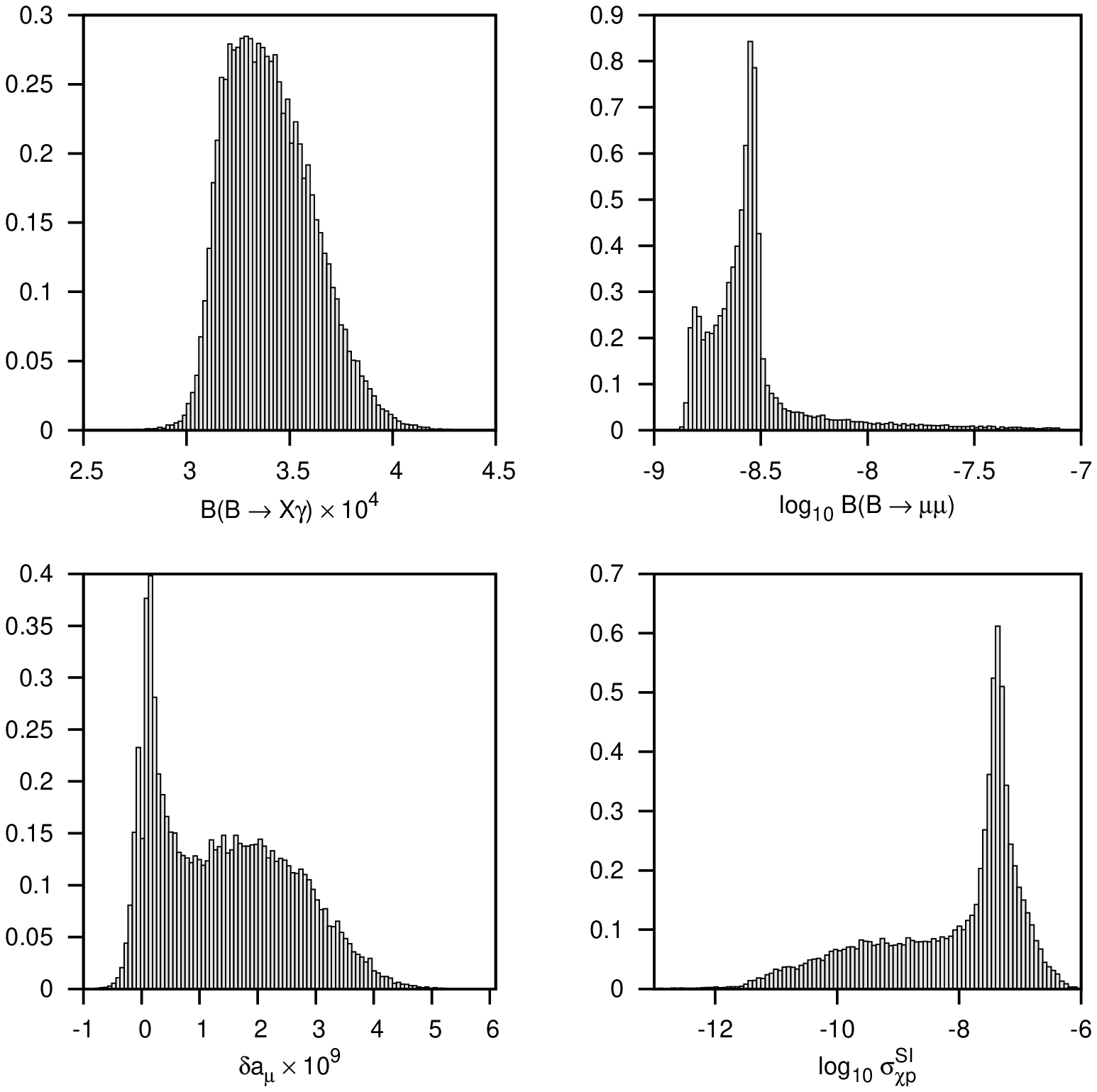}
\caption{Distributions for $\bsg,\bsmu,\amu,\sip$. The y-axis is in units of 
$10^6$.}
\label{fig:obs}
\end{figure}

 Note that if $\amu$ turns out to be consistent with the SM prediction, as indicated by some preliminary 
results~\cite{Davier:gmuon},  the distributions for  $M_2,\msl$ would shift to larger values. 
The dependence of the B-observables on some of the most relevant parameters are displayed in Fig.~\ref{fig:2d}c-i,
in particular  the enhancement with the light pseudoscalar mass for $\bsmu,\bsg,\btau$. 
In fact  a heavy pseudoscalar leaves little prospects for observing large deviations
in B-observables,  $M_A>1$~TeV implies $\bsmu<5\times 10^{-9}$ and  $\btau>0.70$. Similarly
small values of $\tan\beta$, for example $\tan\beta<20$ implies $\bsmu<2.5\times 10^{-9}$ and  $\btau>0.85$. 
The potential enhancement at large values of $\tan\beta$ is illustrated only for $\bsmu$.  
The largest values  for $\bsmu$ are also found  at low values of  $M_2$, they are therefore correlated with large deviations in $\amu$
~\cite{Dedes:2001fv}.

\begin{figure}[ht]
\centering
\includegraphics[width=14cm,angle=0]{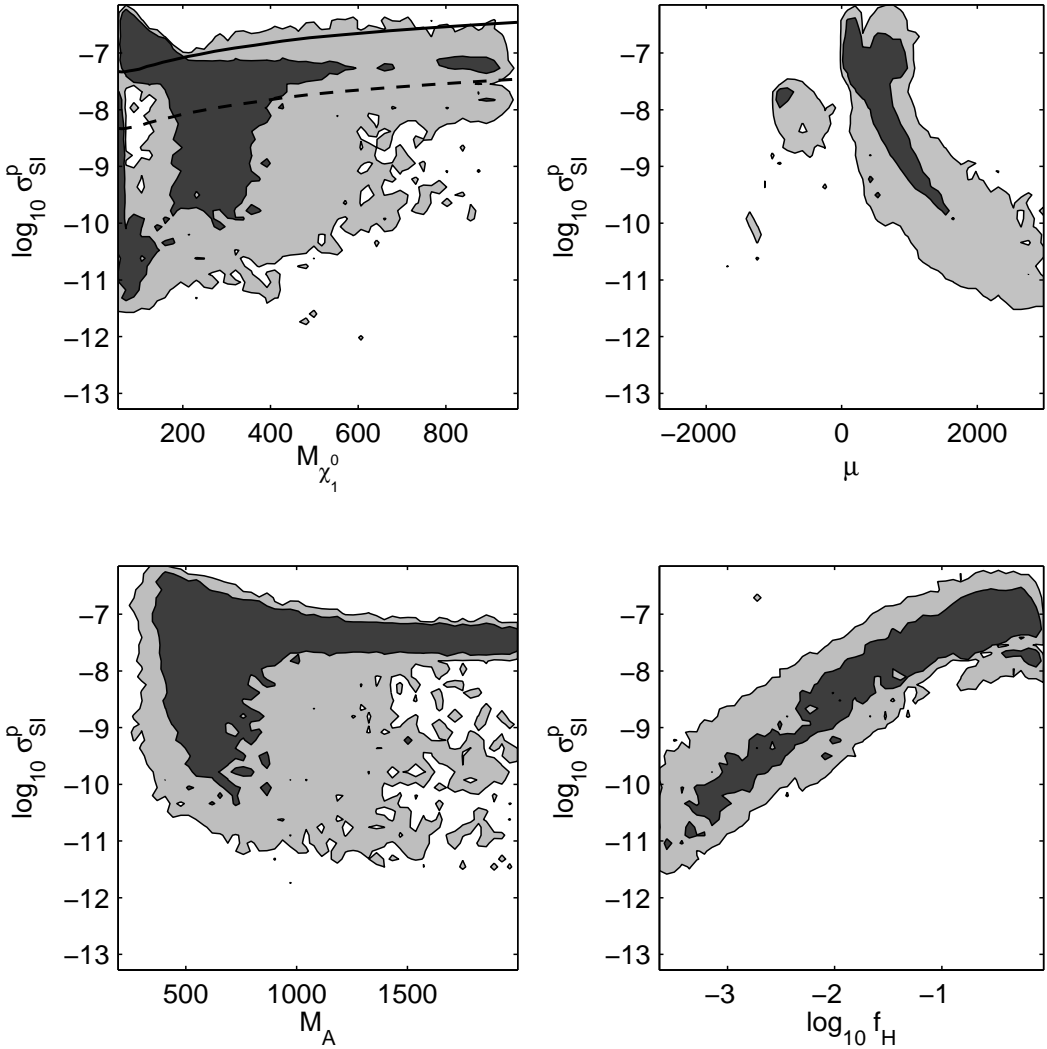}
\caption{Predictions for elastic scattering cross sections as a function of $\mneut$, $\mu$, $M_A$ and $f_H$.
Default values of micrOMEGAs for the quark content of the nucleon are taken. On
the top-left plot the CDMS-II exclusion is shown by a solid line, while the 
dashed line show the factor 10 improved limits.}
\label{fig:SI}
\end{figure}

The predictions for the neutralino nucleon elastic scattering, $\sip$,  vary 
over 5 orders of magnitude. The distribution peaks at values near the experimental upper limit, see Fig.~\ref{fig:obs} and can even exceed 
the present limit. The LSP mass dependent CDMS limit is displayed in Fig.~\ref{fig:SI}a ~\cite{Ahmed:2008eu}. 
%RKS: Could you re-write the sentence below. -ok gb
The light Higgs exchange in general dominate SI interactions. Therefore large cross sections
 are expected when the LSP has a large coupling to the light Higgs, this means some higgsino component, Fig.~\ref{fig:SI}b,d.
%Orders of magnitude differences are found for the bino LSP as compared to the dominantly higgsino one. 
For example  a LSP with a higgsino component $f_H>0.2$  necessarily  implies $\sip>3.2\times 10^{-9}$~pb.
  These values will be probed in the near future with  for example
 Xenon100~\cite{Aprile:2009yh}. On the other hand a pure bino LSP has a cross section at least two orders of magnitude below the experimental limit. 
 The second Higgs scalar as well as squarks can also contribute significantly when they 
 have a mass comparable to the LSP. There is however no direct correlation between the direct detection rate and $M_A$.  
Note that the enhancement of the heavy Higgs contribution relative to the light
 Higgs expected at large values of $\tan\beta$ is tamed because the pseudoscalar is always much heavier than the light Higgs
 (recall that $M_A>370$~GeV).

\newpage

\subsection{Complementarity of different LHC searches and direct dark matter searches} 

From the above discussion, it is clear that B-physics observables, $\amu$, direct detection as well as direct particle searches at LHC 
are sensitive to different sectors of the MSSM. 
Direct searches at LHC are especially powerful to probe the coloured sector while heavy Higgs searches probe 
the large $\tan\beta$. Direct detection probes best the higgsino LSP.

To illustrate quantitatively the complementarity of the various new physics signals at LHC and in direct detection we compute the fraction of models in the allowed parameter space
that lead to a signal in 4 different channels: direct searches of sparticles at LHC, searches for heavy Higgs at LHC, deviation in $\bsmu$ and DM direct 
detection.  Here we choose $\bsmu$ as a representative observable from the B-sector to simplify the discussion. The same exercice with another
B-observables would lead to a similar conclusion.
The signal in each channel is define as follows. 
We assume that coloured sparticles are within reach at LHC-14TeV if their mass is below 2~TeV. This corresponds roughly to the reach for gluinos when 
squarks are heavy and  ${\cal L}=60fb^{-1}$~\cite{atlas_tdr,CMS_tdr}. This value for the squarks reach is  conservative but we use it for simplicity, recall that with a high luminosity, ${\cal L}=300fb^{-1}$, coloured sparticles can be discovered up to 3~TeV. 
To define the  discovery reach for the heavy Higgs we use the results of the CMS study which gives the discovery region 
 in the $\tan\beta-M_A$ plane   for ${\cal L}=30 fb^{-1}$~\cite{CMS_tdr}. This is based on the associated production of a Higgs with
 b-quarks, $gg,qq\rightarrow b\bar{b}h$ 
with the Higgs decaying into tau pairs. In the MSSM, this process  is accessible only at large values of $\tb$ 
because of the $\tb$-enhanced couplings of the heavy Higgs to $b\bar{b},\tau\bar{\tau}$ pairs.
For example the pseudoscalar mass 
can be probed up to $M_A<800$~GeV for $\tan\beta$=50. 

For the process $B_s\to \mu^+\mu^-$ we assume the value for the branching $\bsmu<5\times 10^{-9}$ which will be probed even with a low luminosity at the LHC.
Finally for the elastic scattering cross section  our definition of a signal corresponds to one order of magnitude 
improvement over the current CDMS limit~\cite{Ahmed:2008eu}. With various experiments increasing their sensitivities regularly, this level should be reached in the near future~\cite{Aprile:2009yh}.

\begin{table}[!ht]
\caption{Prospects for Higgs searches, direct detection and $\bsmu$ in scenarios with heavy coloured sparticles,
$m_{\tilde{q}},m_{\gluino}>2$~TeV} 
\label{tab:no}
\begin{center}
\begin{tabular}{|c|c|cc|}\hline
Heavy Higgs at LHC &  $\sip$(~pb)  & $\bsmu$&\\ 
         &      &  $>5.\times 10^{-9}$ & $< 5.\times 10^{-9}$\\ \hline
            
Yes     &  20.6\% ($>10^{-9}$) & 5.6\% &15\%\\
24.2\%  &  3.6\%($<10^{-9}$)   & 1.7\% & 1.9\% \\\hline
No      & 54.9\%($>10^{-9}$)   & 0.3\%& 54.6\% \\
75.8\%  & 20.9\%($<10^{-9}$)   & 0.7\%& 20.2\% \\\hline
\end{tabular}
\end{center}
\end{table}

% Even this high luminosity will leave some of the allowed parameter space unexplored.  
%We present the results for the fraction of allowed models that lead to a signal in each mode in Table~\ref{tab:no}. 
%We display  separately  the cases where sparticles are or are not within the reach of the LHC.
%A glance at the tables shows that the results are quite similar so we will discuss in more details only t
First we consider the case where sparticles are too heavy to be 
accessible at LHC, in these scenarios it is clearly essential to consider alternative signals of new physics 
and/or dark matter. In fact a large number of the allowed scenarios 
(almost 29\%)  predict a heavy coloured spectrum. We will refer to this as set A in the next section.  
This is easily explained by the fact that no observable 
(except weakly for $\amu$) requires a supersymmetric contribution as long as the DM can give the proper relic abundance. 
%Clearly if sparticle searches are unsuccessful at LHC,  
The best alternative at LHC for a sign of physics beyond the standard model would be the search for a heavy Higgs 
\footnote{Here we have not included the possibility of discovering SUSY particles in Higgs decays} with nearly 24.2\% of scenarios predicting  a
 signal  while few deviations from  the SM in $\bsmu$ are expected (only in 5.6\% of the total number of scenarios in set A). 
Furthermore signals 
 in B-physics are mostly found in scenarios  for a light pseudoscalar Higgs and very large values of $\tan\beta$ where one also has a heavy Higgs signal. 
%Conversely for the models with no Higgs signal at LHC, 
% which either means small$\tan\beta$ or large $M_A$, no signal in $\bsmu$ is expected. 
%The prospects for signals in direct detection is also displayed in Table~\ref{tab:no}. T
The complementarity between the direct detection and collider 
searches is clearly evident in Table~\ref{tab:no}. Nearly half (54.9\%) the models lead to a signal only in direct detection experiments
and about 20.6\% of models predict  a signal in both the Higgs and the direct detection. 
Conversely  if no signal in direct detection is observed in the near future the model would be severely constrained 
(with 24.5\% of scenarios remaining).   Of course this statement should be modulated by the fact that there are large theoretical and astrophysical  uncertainties in the prediction of the neutralino proton scattering cross section. 
Combining  all the channels  we are left with a small fraction of models  with no signal in SUSY or  Higgs 
searches at colliders, B physics or direct detection, the fraction corresponds to 20.2\% of the subset with heavy sparticles which means  about 
6\% of the sample of allowed scenarios. We have also checked that there are no other SUSY signals at LHC in these scenarios with heavy squarks and gluinos. We have computed the trilepton signal for direct neutralino/chargino production,    
$pp\rightarrow \tilde{\chi}_i^0\tilde{\chi}_1^+\ra 3l+E_{T}^{miss}$ and found a maximum cross section of a few fb at $\sqrt{s}=14$~TeV. 
The reasons for the small cross section are the heavy squarks and the heavy  neutralinos and charginos, the
(97.5\%C.L.) lower limits are $\mneut,\mneutt,\mneuth>285,366,371>$~GeV and  $\mchar>354$~GeV. 
The direct production of sleptons is also below the LHC reach for $30 fb^{-1}$ since the lower limit on slepton masses are $\msel,\mser>300$~GeV.

\begin{table}[!ht]
\caption{Prospects for Higgs searches, direct detection and $\bsmu$ in scenarios with light coloured sparticles,
$m_{\tilde{q}}< 2$~TeV and $m_{\gluino}<2$~TeV } 
\label{tab:comp}
\begin{center}
\begin{tabular}{|c|c|cc|}\hline
Heavy Higgs at LHC &  $\sip$(~pb)  & $\bsmu$&\\ 
         &      &  $>5.\times 10^{-9}$ & $< 5.\times 10^{-9}$\\ \hline
            
         Yes  &  21.9\% ($>10^{-9}$)  & 7.2\% &14.7\%\\
      24.5\%  &  2.7\%($<10^{-9}$)    & 1.6\% & 1.1\% \\\hline
   No         & 53.7\%($>10^{-9}$)    & 0.6\%& 53.1\% \\
75.5\%        & 21.8\%($<10^{-9}$)    & 1.1\%& 20.7\% \\\hline
\end{tabular}
\end{center}
\end{table}

The results for the case where sparticles are accessible at LHC leads to roughly the same conclusion, see
 Table~\ref{tab:comp}. One  difference
 is a small increase in the fraction of models that predict an observable deviation in $\bsmu$. This is particularly true 
 if  squarks below the 2TeV scale are present. In the case where a heavy Higgs and squarks are accessible at LHC, more
  than 35\% of the models also predict $\bsmu> 5.\times 10^{-9}$.

\subsection{Impact of improved sensitivity}
Rather than just looking at the discovery potential we also examine the impact of a signal in 
the near future in direct detection or in  $\bsmu$. 
First let us comment on the impact of the present CDMS limit on direct detection. Recall that because of the astrophysical and nuclear uncertainties
we have not used this observable in the fit. We applied this constraint a posteriori on the selected models and looked at the impact on
the SUSY spectrum. The most noticeable effect of this constraint is that it removes some models with
$65<\mneut<170$~GeV  especially when those  models have a large value for the higgsino fraction $f_H$. Models with $\mneut\approx 50$~GeV that have a
small higgsino fraction are not affected. Furthermore some of the models with low values of $M_A$ and large $\tb$ are disfavoured. 
To have an indication of the impact of a signal in direct detection we then considered the case of a signal $\sip=1.\times 10^{-8}$~pb  
%RKS : I have cut the total data for \sip between 1e-8/3 and 3e-8. What do you
% mean by same mass dependece as CDMS ?? okgb
and allowed a factor 3  theoretical uncertainty. With such a measurement one could constrain the
allowed parameter space although no specific correlation with other observables are observed, only a lower bound on the LSP higgsino
fraction is found. 

We have also considered the impact of a signal at the Tevatron in $\bsmu$, say     
 $\bsmu>1.8 \times 10^{-8}$ which represents the ultimate reach of Tevatron~\cite{moriond_talk}.
We found that in this case a lower bound on $\tan\beta>32$ with 97.5\%C.L could be extracted (though a tail extends till $\tan\beta\sim 18$) and furthermore
 that a  heavy Higgs signal at LHC would  almost be guaranteed (in 94.6\% of the scenarios).
Finally the models with a signal within the Tevatron reach in $\bsmu$ all predict a small cross section in the trilepton channel 
($\sigma_{lll}^{\rm TeV} < 0.1$~fb) even before applying cuts. This points towards a complementarity between the two channels 
as was found in the MSSM~\cite{Dedes:2002zx}, although 
as we have mentionned before only a small fraction of the MSSM-UG models are within the Tevatron reach for the trileptons.

\section{LHC forecasts}

At last we examine in more details  the SUSY signatures at the LHC. 
For this we  split the  scenarios allowed at 95\% C.L. in different sets according
to which sparticles are within the reach of the LHC. We have already mentionned that  a  significant fraction (28.7\%) of our scenarios
predict a coloured spectrum that is just too heavy (set A). 
Another large fraction (30.5\% ) corresponds to the case where the gluino is the only coloured sparticle that could be discovered at LHC (set B)
 while a squark is also within reach in 22.0\% (set C). In 19\% of our scenarios (set D) only squarks are accessible. 
The main difference from  the CNMSSM is that in the latter the squark masses receive an important contribution from  $M_3$ due to the renormalization group
running so the cases where only a
 gluino is accessible at LHC are confined to the focus point region. 
The precise fraction of models that have sparticles within reach of the LHC do depend on the range on the parameters used in the scan.
 In particular it is clear that a wider range for the squark masses would have lead to a larger fraction of models with squarks too heavy
  to be produced at the LHC.

For two sets of scenarios we then examine the main branching fractions of sparticles concentrating on squarks, 
gluinos and on the neutralino/charginos produced in their decay.  Our goal is to point out the different decay channels available.
We do not attempt to analyse the feasibility of extracting signals over background in our scenarios, this  is beyond the scope of this paper.   

Note that in some scenarios, the chargino/neutralino can be produced directly
leading to a  signal in the trilepton channel,     
$p\bar{p}\rightarrow \tilde{\chi}_i^0\tilde{\chi}_1^+\ra 3l+E_{T}^{miss}$. 
We do not here analyse direct production specifically
but we have checked that the Tevatron only weakly constrained our allowed scenario, only about 1 per-mil
of our models have a trilepton cross section $\sigma(3l+E_{T}^{miss})>0.0.08$~pb before applying cuts. 

\subsection{Scenarios where $m_{\gluino}<2$TeV, $M_{\tilde q}>2$~TeV.}

In the MSSM-UG  the scenarios  where only gluinos are accessible at LHC represent a large fraction of the scenarios
 that provide a good fit to the data. To examine the characteristic decay chains in these scenarios, 
we have, for each point in our MCMC chain, computed the
  branching ratios of gluinos into neutralino/charginos as well as the 2- and 3-body branching ratios of $\neutt$ and $\neuth$ with \micromegas.
  The decay chains involving a chargino can be dominant but 
  we concentrate on the heavy neutralinos because the leptonic decay mode  
provide a clear signature and the kinematic endpoint in the lepton invariant mass distribution further allow for a determination of the mass difference 
of the heavy neutralino with the LSP. 
  
\begin{figure}[!ht]
\centering
\includegraphics[width=6cm,angle=-90]{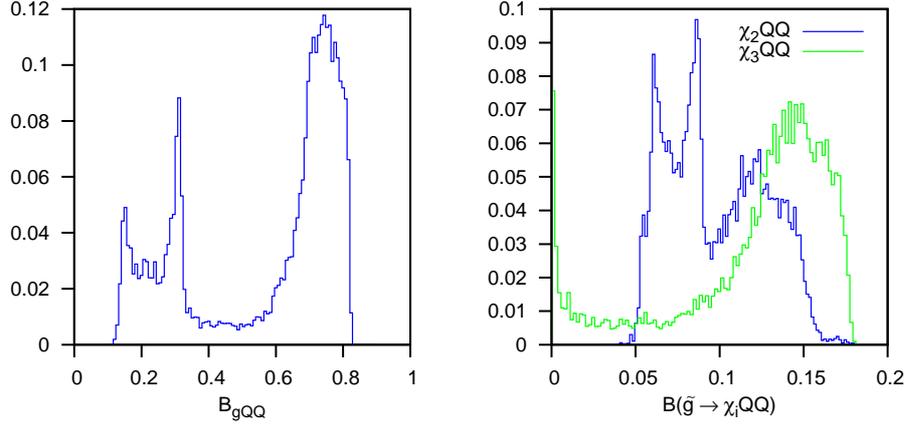}
\caption{Distribution of the gluino branching fractions in third generation
quarks a)$B_{gQQ}$ b) $Br(\gluino\rightarrow \neuti QQ)$. 
There is a scale factor of $10^6$ on the y-axis.}
\label{fig:gluino_br}
\end{figure}

\begin{figure}[!ht]
\centering
\epsfig{file=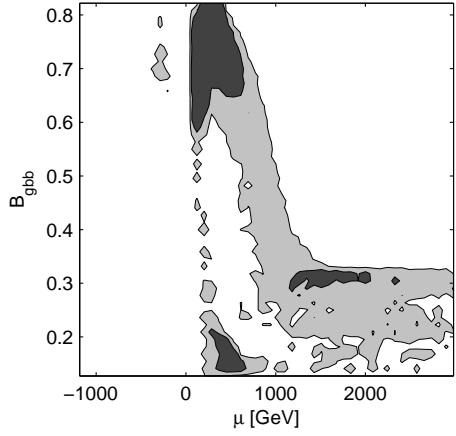,width=6.0cm}
\caption{$B_{gQQ}$ as a fuction of $\mu$ parameter.}
\label{fig:new}
\end{figure}

In these scenarios  the squarks are heavy so the gluino decays only via 3-body, $\gluino\ra \chargi f f'$ or 
$\gluino\ra \neuti f \bar{f}$ leading to final states with  many jets as well as missing energy from the LSP. 
The branching fraction into third generation quarks is particularly relevant since
it has been shown ~\cite{Mercadante:2005vx,Kadala:2008uy}
%check also in newer paper ~\cite{Kadala:2008uy}
 that requiring tagged b-quark jets in the final state helped reduce the SM background  and thus could extend the LHC reach in this channel. 
In our sample scenarios, the total branching fraction for 3-body decays of gluinos into third generation quarks ($t$ and $b$) 
$B(\tilde{g}\rightarrow \chi QQ')=B_{gQQ}$ varies over a wide range, from 
$0.1-0.8$ and features two peaks around 25\% and 70\%, see 
Fig.~\ref{fig:gluino_br}a.\footnote{For reference recall that in a 
typical focus point scenario the branching ratios of gluinos 
into third generation quarks reach 72\% ~\cite{DeSanctis:2007td}.}
The $\neutt,\neuth$ channels each contribute below 20\%, see Fig.~\ref{fig:gluino_br}b.
The larger branching fractions $B_{gQQ}$ are found  for
low values of $\mu$, see Fig.~\ref{fig:new},  
this is mainly because in this case all  charginos and neutralinos can be produced in gluino decays and contribute 
significantly (especially charginos). 
%In particular the heaviest chargino and neutralino contribute significantly. 
The branching fraction $B_{gQQ}$ also increases with $\tan\beta$, this is because of the enhanced coupling of the higgsino to $\tilde{b}b$.

\begin{figure}[!ht]
\centering
\includegraphics[width=5cm,angle=-90]{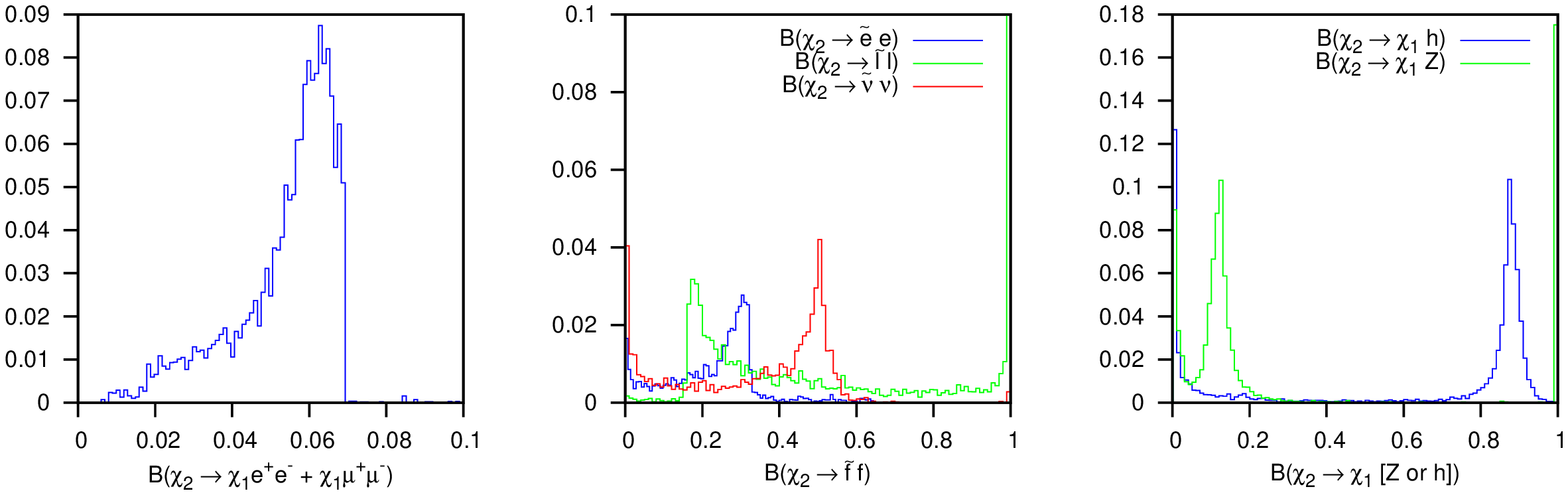}
\caption{Distribution of $\neutt$ branching fractions for the channels  a) $\neuto e^+e^- (\mu^+\mu^-)$ 
b) $\tilde{l}l$ c) $\neuto Z,\neuto h$. There is a scale factor of $10^6$ on the y-axis.}
\label{fig:neut2_br}
\end{figure}

\begin{figure}[!ht]
\centering
\includegraphics[width=5cm,angle=-90]{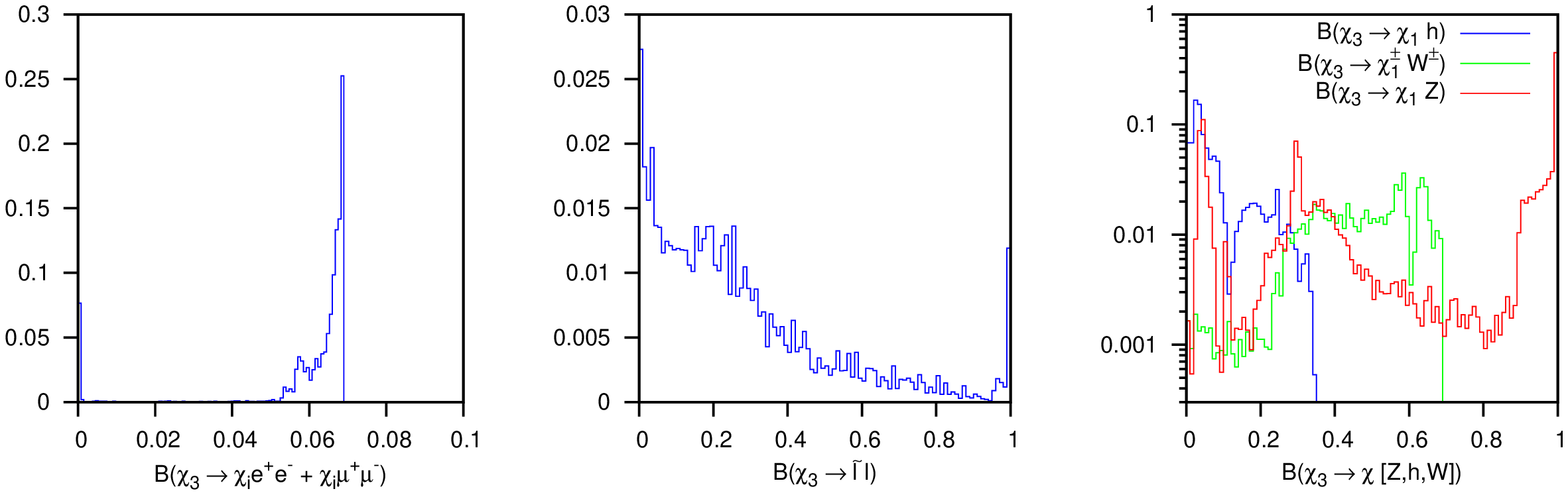}
\caption{a) Distribution of $\neuth$ branching fractions for the channels  a) $\neuti e^+e^-,\mu^+\mu^-$ 
b) $\tilde{l}l$ c) $\neuto Z,\neuto h,\charg W$. There is a scale factor of $10^6$ on the y-axis.}
\label{fig:neut3_br}
\end{figure}

\begin{table}[!ht]
\caption{Dominant decay mode of $\neutt$ and $\neuth$  in models with $m_{\gluino}<2$~TeV and $\msq>2$~TeV and fraction of allowed models corresponding to each dominant decay mode.} 
\label{tab:gluino}
\begin{center}
\begin{tabular}{|c|c|c|}\hline
  $\neutt$     & $\neuth$    & Fraction of models  \\ \hline
 $\neuto f\bar{f}$   & $\tilde{\chi}_i^0 f\bar{f}$  & 35.2\% \\\hline
	          & $\chi_1^\pm W^\mp$ & 4.2\% \\  
  $\neuto f\bar{f}$      &  $\chi_i Z,\chi_i h$ & 8.6\%  \\  
       		  &  $\tilde{l}l$ & 0.2\%  \\  \hline
      $\neuto Z  $   & $\chi_1^\pm W^\mp$ & 6.0\%  \\  
       $\neuto h  $ & $\chi_i Z$  (or $\chi_1^\pm W^\mp$ ) & 14.4\%  \\  
         $\tilde{l}l$  & $\tilde{l}l$    & 4.6\%   \\  
                     & $\chi_1^\pm W^\mp$    & 6.1\%  \\ \hline 
  $\neuto f\bar{f}$    & - & 3.8\%  \\ \hline
  $\neuto Z$    & - & 2.8\%  \\  
   $\neuto h$    & - & 6.7\%  \\ 
    $\tilde{l}l$    & - & 5.0\%  \\ \hline
	            
 \end{tabular}
\end{center}
\end{table}

Next we have computed the decay modes of the $\neutt$ and $\neuth$. The $\neutt$ decays dominantly via 3-body, the branching fraction
 of neutralinos  into each fermion pair
 is typical of the Z decay (around 3\% for each leptonic mode), see Fig.~\ref{fig:neut2_br}. In some scenarios the  
 two-body leptonic channels are accessible, in this case the $\neutt$ either decay exclusively into the ligthest slepton (the stau)
 or into all sleptons including the invisible mode into sneutrinos, see Fig.~\ref{fig:neut2_br}b. 
 Finally the decay channels into a gauge or Higgs boson and the LSP dominate for sufficient mass splitting between the neutralinos, 
 the dominant mode is $\neutt\ra \neuto h$ unless only the Z is kinematically accessible. 
 For  $\neuth$ a larger fraction of scenarios have 2-body decay modes, in particular the decays into neutral and charged gauge bosons,
 $\neuth\ra \lsp Z, {\rm or} ~\charg W$. Decays into sleptons are as before confined to the scenarios with light sleptons,
 Fig.~\ref{fig:neut3_br}.
   
Finally we have compiled the fraction of models (among set B only) that feature each dominant decay mode, 
the results are presented in Table~\ref{tab:gluino}. 
The largest sample (35\%) corresponds to both neutralinos decaying into 3-body channels.  
In these scenarios  $\mu<350$~GeV, $\mu - M_1$ is small and the LSP has a significant higgsino component so that the direct 
detection rate is large ($\sip> 7\times 10^{-9}$~pb).
The scenarios where $\neuth$ instead has 2-body decays also have in general a LSP with an important higgsino component
and a large $\sip$ except when $\neuth\rightarrow \charg W^-$ is the dominant channel. Then
 we found predominantly $\mneut\approx 60$~GeV with a bino LSP annihilating  efficiently through a light Higgs resonance.
 %Because of the resonance effect in neutralino annihilation, there is no need for a large higgsino fraction so that  
 %$\sip$ can be as low as  $3\times 10^{-10}$~pb whereas for the $Z$ or sleptons mode much larger rates are expected, $\sip>10^-8$~pb.

A large number of scenarios (31\% of set B) correspond to both $\neutt$ and $\neuth$ having 2-body decay modes.  
The case where the dominant decay mode is $\neutt\rightarrow \neuto Z$ correspond to a small value for  $\mu-M_1$, 
this means a dominantly higgsino LSP. In general though the dominant modes are $\neutt\ra\neuto h$ and $\neuth\rightarrow \neuto Z$ 
unless sleptons are light. In nearly 19\% of sceanrios B, $\mu$ is large and the $\neuth$ which is higgsino cannot be produced
  in gluino decays.  The $\neutt$ decays predominantly via 2-body channels,  $\tilde{l}l,\chi_1 h $ or $\chi_1 Z$ or
into 3-body $\neutt\ra \neuto f\bar{f}$ with the leptonic mode offering a robust signature. 
Typically in  these scenarios the elastic scattering cross sections is small 
apart from the cases where the heavy Higgs is light enough to contribute significantly.

In conclusion, although the focus point like scenario with gluinos decay with large fraction into $b$ quarks and leptonic 3 body decays of 
$\neutt,\neuth$
constitute the largest sample, the channels with two body decays of neutralinos are also very important. 
Among these cases where  leptonic decays of $\neutt$ dominate constitute 16\% of our sample while those where the only
final states include $Z(W)$ or $h$ constitute about 30\% of our sample.  
Note that the gluino only scenario could be quite challenging for the LHC especially if the gluino is near 2TeV. 
We have checked that  the scenarios that
would be accessible with a lower luminosity, say   $m_{\gluino}<900$~GeV for ${\cal L}=30{\rm fb}^{-1}$, 
share the main features we have just described.  Obviously because of the
lighter spectrum  a smaller fraction of scenarios have $\neuth$  accessible in gluinos decays. 
Furthermore the fraction of scenarios with 2-body decays  is smaller with for example hardly any scenario where both
$\neutt$ and $\neuth$ decay via 2-body.

\subsection{Scenarios where $\msq<2$~TeV, $m_{\gluino}>2$~TeV }

In the MSSM-UG a small fraction of allowed scenarios (set D) predicts that only squarks are lighter than  2TeV.   
In this case, the main contribution to SUSY production is direct production of squark pairs. 
In the MSSM-UG, the masses of sleptons and squarks are not
correlated ,so sleptons are not always light enough to be produced in neutralino decays.
The conventional  SUSY searches which involve  a decay chain with a neutralino 
$\tilde{q}\ra\neutt j\ra \tilde{l}l j\ra \neuto \tilde{l} l j$ will  be superseded by the two-body decays,
$\neutt\ra \neuto Z,\neuto h$ or $\charg W$ or if kinematically forbidden by  3-body decays.
%or with   a chargino $\tilde{q}\ra\charg j\ra  W \neuto j$.  
%When this is the case preferred two-body decays will rather be  . 
%If these channels are kinematically forbidden then 3 body decays can occur. 
%For the chargino, the main 2 body channel is into $\charg\ra \neuto W$  unless sleptons are light.

We have computed the branching fractions  for squarks, taking as an example the $\tilde{u}_L$ and $\tilde{u}_R$. 
We have then looked for the dominant mode into either a chargino, a heavy neutralino or the direct decay into the LSP. 
In each case we have also
examined the decay mode of the neutralinos and charginos that occur in the squark decays.
The higgsino fraction of the LSP is an important factor that determines the dominant decay mode. 
 We therefore analyse the dominant decay modes separately for the case of a bino, mixed or higgsino LSP. 
 Note that a mixed LSP constitutes  more than half of our sample of  scenarios.

The right-handed squark  decays mainly into the bino component which means that the dominant decay is in general 
$\tilde{u}_R\ra u\neuto$, leading to only jets and missing energy. The decay $\tilde{u}_R\ra u\neuth$ can occur only 
when the LSP is a  higgsino ($f_H\gsim 50\%$) so that the $\neuto$ channel is suppressed.
In this case however the heavier neutralino decay mostly via  3 body decays into the lightest chargino or neutralino 
+ jets leading again to signatures with jets and missing energy.

\begin{table}[!ht]
\caption{Dominant decay  chains for $\ul$ for a bino, mixed or higsino LSP.} 
\label{tab:ul}
\begin{center}
\begin{tabular}{|cl|c|c|c|}\hline   
 & &  bino  & mixed  & higgsino\\ 
 & &  $f_H\le 0.01$  & $0.01<f_H\le 0.5$ & $f_H>0.5$\\ \hline
$\ul\ra\neuto~u$  & & 7.7\% & 2.5\% & $ - $      \\\hline
  $\ul\ra\neuth~u$    & $\neuth\ra \sl l$             & - & -  &  0.3\%  \\
                   & $\neuth\ra \charg W $             & - & -  &  0.3\%  \\
 		   & $\neuth\ra \charg f' \bar{f}$ & - & 2.0\%  &  7.4\%  \\\hline
$\ul \ra \charg d$                   & $\charg\ra \neuto W$& 12.4\%& - & -  \\
		   & $\charg\ra  \sl l' $& 8.9\%& - & - \\\hline
 $\ul \ra \chargt d$  & $\chargt\ra \neuti W$          & 2.7\% & 37.2\%  &2.5\% \\
		   & $\chargt\ra \charg h$          & -     & 1.0\%  & 0.6\% \\
		   & $\chargt\ra  \sl l' $          & 1.1\% & 12.5\%  & 0.6\% \\\hline
Fraction of sample && 32.8\%&55.3\% & 11.9\%\\\hline
\end{tabular}
\end{center}
\label{tab:squark}
\end{table}

The left-handed quark which couples strongly to the wino and/or higgsino component has a wide variety of decay modes. 
The frequency of each dominant decay chain are displayed in Table~\ref{tab:squark} for each LSP configuration. 
For the bino LSP the dominant mode is usually $\tilde{u}_L\ra d\charg$ with typical branching fractions around 60\%.
The chargino will decay either into $\neuto W$ or $\sl l'$ when light sleptons are present. The subdominant mode in those scenario is 
$\tilde{u}_L\ra u\neutt$ with $\neutt\ra \sl l,\neuto Z, \neuto h$. The decay chains are similar to those of the CMSSM. 
 In some cases the second chargino, a mixed higgsino/wino, is kinematically accessible and the dominant mode will be 
  $\tilde{u}_L\ra d\chargt$ with subdominant decays into $\neutf u$ and $\charg d$. $\chargt$ will decay preferentially
   into $\neutt W$ or in other neutralinos as well as into $\sl l$. The Higgs can be produced in either $\chargt\ra \charg h$ 
   or further in $\neutt\ra \neuto h$. A fraction of models (7.7\%) feature the dominant decay into the LSP $\tilde{u}_L\ra u\neuto$.
    Because the squark $\tilde{q}_L$ has a suppressed rate to the bino, this channel is dominant only when other two-body channels are kinematically
     forbidden.

For a mixed LSP ($0.01<f_H\le 0.5$) the relative importance of the various decay channels shifts. The decay 
$\tilde{u}_L\ra u\neuto$ is dominant in less than 3\% of the cases although because of the higgsino component of the LSP this can occur even when heavier neutralinos are kinematically accesssible. By far the most frequent dominant decay is $\tilde{u}_L\ra d\chargt$ with significant branching fractions in 
$ u\neuth$ or $d\charg$. The heavier chargino always has two-body decay modes, $\chargt\ra \neuti W$ 
(preferably $\neutt W$) or $\chargt\ra \charg h$.
The $\neutt,\neuth$ and $\charg$ in turn feature mostly 3-body decays. 
Note that decay modes into Higgs bosons $\chargt\ra \charg h$  can involve even the heavy Higgs bosons. 
As usual when light sleptons are present the decay $\chargt\ra \sl l' $ can be dominant.  
%The fraction of models for each dominant decay channel are listed in Table, for the $\chargt\ra \neuti W$ channel a sum over the neutralino states was performed. Since in most cases there are significant  branching fraction into at least two neutralino states, the fraction of models where an individual mode is dominant is much smaller. When comparing with individual channel the branching fraction into the light Higgs is often the dominant one.  

For the higgsino LSP ($f_H>0.5$), the dominant mode is  either $\neuth u$ or $\chargt d$ with some contributions from $\chargt d$
and $\neutf u$. The $\chargt$ channel has similar decay chains as the mixed LSP  except that the dominant mode is usually $\chargt\ra \neutt W$ rather than channels involving Higgses.
The $\neuth$ can  in a few cases  decay via two-body, $\charg W$  or $\sl l$, but in most cases it decays  via three-body dominantly into
$\neuth\ra \charg ff'$.  These decays mainly give signatures into jets and missing energy.  The $\charg$ produced in squark or neutralino 
decays will also decay via three body final states. 

In summary  $\tilde{q}_L$ decays dominantly into heavy charginos with further decay chains involving
 other chargino/neutralino states. Decay chains  involving 
slepton production  dominater only in 25\% of scenarios. 
Finally recall   that 
the elastic scattering cross section also differs significantly depending on the nature of the LSP giving an opportunity to
correlate SUSY signals at LHC with those of direct detection.  
 For the bino LSP, $\sip<10^{-9}$~pb while $\sip>10^{-9}(5\times 10^{-8})$~pb for the mixed (higgsino) LSP.

We do not discuss in detail the case where both squarks and gluinos
 are below 2TeV. The decay chains can be rather complicated with 
 the possibility of producing the gluino in squark decay and vice-versa. 
 The case where the gluino is heavier than the squarks features the same
  decay chains for the squarks as the case just discussed.

\section{Conclusion}
Increasing the
number of free  parameters as compared to the  CMSSM model has  opened up the possibilities for supersymmetric scenarios 
that are compatible with all experimental constraints and this even maintaining the
universality of gaugino mass. Although the parameter space of the model is still not very well 
constrained,  we  found that the most favoured models
have a LSP of a few hundred GeV  with a significant higgsino fraction ($>10\%$). 
Contrary to the CMSSM case the higgsino LSP is not fully correlated with a very heavy
squark sector although all our scenarios favour squarks above the TeV scale. A very light pseudoscalar is also disfavoured with
$M_A>370$~GeV, this means that  large deviations from the SM in B-physics observables are expected only in a small fraction of allowed
scenarios. Our favoured scenarios predict few  signals at the Tevatron, 
the pseudoscalar Higgs as well as the coloured sector are too heavy to be accessed by direct searches.
Only very few scenarios have a potentially large enough rate for trilepton searches at the Tevatron.
 The complementarity between future experiments to probe this class of models was emphasized. Even though SUSY or heavy Higgs
 signals are not guaranteed at LHC, the majority of allowed models predict at least one signal either at the LHC (including the flavour sector)
 or in future direct detection experiment. Furthermore the light Higgs is expected to be  around 120GeV with SM-like couplings.
 
We have also explored the various dominant decay chains for gluinos and squarks that could be produced at LHC  in the MSSM-UG 
 as well as for the heavy neutralinos appearing 
in the decays of these coloured sparticles. We found that for models with gluinos accessible at LHC, 
a significant fraction of the 
heavy neutralinos produced decayed dominantly into a gauge or Higgs boson. 
Furthermore states which decayed into sleptons are rarely dominant. We also showed how the preferred squarks decay channels are determined
to a large extent by the neutralino composition. Whether one can exploit these decay chains to determine some properties
of the sparticles remains to be seen. 
In our analysis the relic density measurement plays the dominant role in constraining the model, since the relic density computation
implicitly assumes a standard cosmological scenario, relaxing this requirement affects significantly 
the allowed parameter space of the model.

Finally we comment on the difference between our results and other recent analyses 
done within the framework of the MSSM with 24 parameters, either using
a MCMC likelihood approach or applying $2\sigma$ constraints on each of the observables~\cite{Berger:2008cq,Profumo:2004at}.
First these studies  were done in a more general model than the one we have considered,
with in particular no  universality condition on the gaugino masses.
This means that the LSP can have a significant wino component and therefore is more likely to be at the TeV scale as was found in 
~\cite{AbdusSalam:2009qd} using linear priors. Recall that a TeV scale wino annihilates efficiently into gauge bosons pairs. The analysis of
~\cite{AbdusSalam:2009qd} also emphasizes the prior dependence with a generally much lighter spectrum using log priors.
This is due mostly to the poorly constrained parameter space~\cite{Allanach:2007qk,Trotta:2008bp}.
As in our analysis squarks and sleptons ran over the full range allowed in the scan and the pseudoscalar mass can be very heavy.

The  analysis of ~\cite{Berger:2008cq}  used  a different statistical treatment
but most importantly did not require that the
neutalino explained all the DM in the universe (only an upper bound on $\Omega h^2$ was imposed). This means that a large number of models with
small mass splitting between the LSP and the NLSP appeared in the scan calling for  a careful study of collider limits. 
In our approach
such models are  ruled out since they have $\Omega h^2\ll 0.1$. This analysis further emphasized the light SUSY spectrum in 
their scans
so naturally found preferred LSP mass below the TeV scale.

\section{Acknowledgements}
We thank J. Hamann for many useful  discussions on the MCMC method.
We acknowledge support from the Indo French Center fro Promotion of Advanced
Scientific Research under project number 30004-2.
This work was also supported in part 
by the GDRI-ACPP of CNRS and by the French ANR project {\tt ToolsDMColl}, BLAN07-2-194882.
The work of A.P. was supported by the Russian foundation for Basic Research, grant
RFBR-08-02-00856-a and RFBR-08-02-92499-a.

%=======================================================================

\providecommand{\href}[2]{#2}\begingroup\raggedright\endgroup


\begin{thebibliography}{10}

\bibitem{Angle:2007uj}
{\bf XENON} Collaboration, J.~Angle {\em et al.}, ``{First Results from the
  XENON10 Dark Matter Experiment at the Gran Sasso National Laboratory},''
  \href{http://dx.doi.org/10.1103/PhysRevLett.100.021303}{{\em Phys. Rev.
  Lett.} {\bf 100} (2008)  021303},
\href{http://arxiv.org/abs/0706.0039}{{\tt arXiv:0706.0039 [astro-ph]}}.
%%CITATION = 0706.0039;%%.

\bibitem{Ahmed:2008eu}
{\bf CDMS} Collaboration, Z.~Ahmed {\em et al.}, ``{A Search for WIMPs with the
  First Five-Tower Data from CDMS},''
\href{http://arxiv.org/abs/0802.3530}{{\tt arXiv:0802.3530 [astro-ph]}}.
%%CITATION = 0802.3530;%%.

\bibitem{Adriani:2008zr}
{\bf PAMELA} Collaboration, O.~Adriani {\em et al.}, ``{An anomalous positron
  abundance in cosmic rays with energies 1.5.100 GeV},''
  \href{http://dx.doi.org/10.1038/nature07942}{{\em Nature} {\bf 458} (2009)
  607--609},
\href{http://arxiv.org/abs/0810.4995}{{\tt arXiv:0810.4995 [astro-ph]}}.
%%CITATION = 0810.4995;%%.

\bibitem{Abdo:2009zk}
{\bf The Fermi LAT} Collaboration, A.~A. Abdo {\em et al.}, ``{Measurement of
  the Cosmic Ray e+ plus e- spectrum from 20 GeV to 1 TeV with the Fermi Large
  Area Telescope},''
\href{http://arxiv.org/abs/0905.0025}{{\tt arXiv:0905.0025 [astro-ph.HE]}}.
%%CITATION = 0905.0025;%%.

\bibitem{Aharonian:2008aaa}
{\bf H.E.S.S.} Collaboration, F.~Aharonian {\em et al.}, ``{The energy spectrum
  of cosmic-ray electrons at TeV energies},''
  \href{http://dx.doi.org/10.1103/PhysRevLett.101.261104}{{\em Phys. Rev.
  Lett.} {\bf 101} (2008)  261104},
\href{http://arxiv.org/abs/0811.3894}{{\tt arXiv:0811.3894 [astro-ph]}}.
%%CITATION = 0811.3894;%%.

\bibitem{Adriani:2008zq}
O.~Adriani {\em et al.}, ``{A new measurement of the antiproton-to-proton flux
  ratio up to 100 GeV in the cosmic radiation},''
  \href{http://dx.doi.org/10.1103/PhysRevLett.102.051101}{{\em Phys. Rev.
  Lett.} {\bf 102} (2009)  051101},
\href{http://arxiv.org/abs/0810.4994}{{\tt arXiv:0810.4994 [astro-ph]}}.
%%CITATION = 0810.4994;%%.

\bibitem{Dunkley:2008ie}
{\bf WMAP} Collaboration, J.~Dunkley {\em et al.}, ``{Five-Year Wilkinson
  Microwave Anisotropy Probe (WMAP) Observations: Likelihoods and Parameters
  from the WMAP data},''
  \href{http://dx.doi.org/10.1088/0067-0049/180/2/306}{{\em Astrophys. J.
  Suppl.} {\bf 180} (2009)  306--329},
\href{http://arxiv.org/abs/0803.0586}{{\tt arXiv:0803.0586 [astro-ph]}}.
%%CITATION = 0803.0586;%%.

\bibitem{Spergel:2006hy}
{\bf WMAP} Collaboration, D.~N. Spergel {\em et al.}, ``{Wilkinson Microwave
  Anisotropy Probe (WMAP) three year results: Implications for cosmology},''
  \href{http://dx.doi.org/10.1086/513700}{{\em Astrophys. J. Suppl.} {\bf 170}
  (2007)  377},
\href{http://arxiv.org/abs/astro-ph/0603449}{{\tt arXiv:astro-ph/0603449}}.
%%CITATION = ASTRO-PH/0603449;%%.

\bibitem{Tegmark:2006az}
{\bf SDSS} Collaboration, M.~Tegmark {\em et al.}, ``{Cosmological Constraints
  from the SDSS Luminous Red Galaxies},''
  \href{http://dx.doi.org/10.1103/PhysRevD.74.123507}{{\em Phys. Rev.} {\bf
  D74} (2006)  123507},
\href{http://arxiv.org/abs/astro-ph/0608632}{{\tt arXiv:astro-ph/0608632}}.
%%CITATION = ASTRO-PH/0608632;%%.

\bibitem{Aaltonen:2008my}
{\bf CDF} Collaboration, T.~Aaltonen {\em et al.}, ``{Search for new physics in
  the $\mu\mu+e/\mu+E_{T}\!\!\!\!\!\!\!/\,\,\,\,$ channel with a low-$p_T$
  lepton threshold at the Collider Detector at Fermilab},''
  \href{http://dx.doi.org/10.1103/PhysRevD.79.052004}{{\em Phys. Rev.} {\bf
  D79} (2009)  052004},
\href{http://arxiv.org/abs/0810.3522}{{\tt arXiv:0810.3522 [hep-ex]}}.
%%CITATION = 0810.3522;%%.

\bibitem{Abazov:2009zi}
{\bf D0} Collaboration, V.~M. Abazov {\em et al.}, ``{Search for associated
  production of charginos and neutralinos in the trilepton final state using
  2.3 fb-1 of data},''
\href{http://arxiv.org/abs/0901.0646}{{\tt arXiv:0901.0646 [hep-ex]}}.
%%CITATION = 0901.0646;%%.

\bibitem{Aprile:2009yh}
E.~Aprile, L.~Baudis, and f.~t.~X. Collaboration, ``{Status and Sensitivity
  Projections for the XENON100 Dark Matter Experiment},''
\href{http://arxiv.org/abs/0902.4253}{{\tt arXiv:0902.4253 [astro-ph.IM]}}.
%%CITATION = 0902.4253;%%.

\bibitem{Bruch:2007zz}
{\bf CDMS} Collaboration, T.~Bruch, ``{Status and future of the CDMS
  experiment: CDMS-II to SuperCDMS},''
\href{http://dx.doi.org/10.1063/1.2823758}{{\em AIP Conf. Proc.} {\bf 957}
  (2007)  193--196}.
%%CITATION = APCPC,957,193;%%.

\bibitem{Moiseev:2008zz}
{\bf GLAST LAT} Collaboration, A.~A. Moiseev, ``{Gamma-ray Large Area Space
  Telescope: Mission overview},''
\href{http://dx.doi.org/10.1016/j.nima.2008.01.005}{{\em Nucl. Instrum. Meth.}
  {\bf A588} (2008)  41--47}.
%%CITATION = NUIMA,A588,41;%%.

\bibitem{Mocchiutti:2009sj}
E.~Mocchiutti {\em et al.}, ``{The PAMELA Space Experiment},''
\href{http://arxiv.org/abs/0905.2551}{{\tt arXiv:0905.2551 [astro-ph.HE]}}.
%%CITATION = 0905.2551;%%.

\bibitem{atlas_tdr}
``{ATLAS detector and physics performance. Technical design report. Vol. 2},''.
  CERN-LHCC-99-15.

\bibitem{CMS_tdr}
{\bf CMS} Collaboration, G.~L. Bayatian {\em et al.}, ``{CMS technical design
  report, volume II: Physics performance},''
\href{http://dx.doi.org/10.1088/0954-3899/34/6/S01}{{\em J. Phys.} {\bf G34}
  (2007)  995--1579}.
%%CITATION = JPHGB,G34,995;%%.

\bibitem{Buchalla:2008jp}
M.~Artuso {\em et al.}, ``{$B$, $D$ and $K$ decays},''
  \href{http://dx.doi.org/10.1140/epjc/s10052-008-0716-1}{{\em Eur. Phys. J.}
  {\bf C57} (2008)  309--492},
\href{http://arxiv.org/abs/0801.1833}{{\tt arXiv:0801.1833 [hep-ph]}}.
%%CITATION = 0801.1833;%%.

\bibitem{Ellis:2007fu}
J.~R. Ellis, S.~Heinemeyer, K.~A. Olive, A.~M. Weber, and G.~Weiglein, ``{The
  Supersymmetric Parameter Space in Light of $B^-$ physics Observables and
  Electroweak Precision Data},''
  \href{http://dx.doi.org/10.1088/1126-6708/2007/08/083}{{\em JHEP} {\bf 08}
  (2007)  083},
\href{http://arxiv.org/abs/0706.0652}{{\tt arXiv:0706.0652 [hep-ph]}}.
%%CITATION = 0706.0652;%%.

\bibitem{Baer:2003yh}
H.~Baer and C.~Balazs, ``{Chi**2 analysis of the minimal supergravity model
  including WMAP, g(mu)-2 and b --> s gamma constraints},''
  \href{http://dx.doi.org/10.1088/1475-7516/2003/05/006}{{\em JCAP} {\bf 0305}
  (2003)  006},
\href{http://arxiv.org/abs/hep-ph/0303114}{{\tt arXiv:hep-ph/0303114}}.
%%CITATION = HEP-PH/0303114;%%.

\bibitem{Belanger:2004ag}
G.~Belanger, F.~Boudjema, A.~Cottrant, A.~Pukhov, and A.~Semenov, ``{WMAP
  constraints on SUGRA models with non-universal gaugino masses and prospects
  for direct detection},''
  \href{http://dx.doi.org/10.1016/j.nuclphysb.2004.11.036}{{\em Nucl. Phys.}
  {\bf B706} (2005)  411--454},
\href{http://arxiv.org/abs/hep-ph/0407218}{{\tt arXiv:hep-ph/0407218}}.
%%CITATION = HEP-PH/0407218;%%.

\bibitem{Baltz:2004aw}
E.~A. Baltz and P.~Gondolo, ``{Markov chain Monte Carlo exploration of minimal
  supergravity with implications for dark matter},''
  \href{http://dx.doi.org/10.1088/1126-6708/2004/10/052}{{\em JHEP} {\bf 10}
  (2004)  052},
\href{http://arxiv.org/abs/hep-ph/0407039}{{\tt arXiv:hep-ph/0407039}}.
%%CITATION = HEP-PH/0407039;%%.

\bibitem{Allanach:2005kz}
B.~C. Allanach and C.~G. Lester, ``{Multi-Dimensional mSUGRA Likelihood
  Maps},'' \href{http://dx.doi.org/10.1103/PhysRevD.73.015013}{{\em Phys. Rev.}
  {\bf D73} (2006)  015013},
\href{http://arxiv.org/abs/hep-ph/0507283}{{\tt arXiv:hep-ph/0507283}}.
%%CITATION = HEP-PH/0507283;%%.

\bibitem{Allanach:2006cc}
B.~C. Allanach, C.~G. Lester, and A.~M. Weber, ``{The Dark Side of mSUGRA},''
  {\em JHEP} {\bf 12} (2006)  065,
\href{http://arxiv.org/abs/hep-ph/0609295}{{\tt arXiv:hep-ph/0609295}}.
%%CITATION = HEP-PH/0609295;%%.

\bibitem{deAustri:2006pe}
R.~R. de~Austri, R.~Trotta, and L.~Roszkowski, ``{A Markov chain Monte Carlo
  analysis of the CMSSM},'' {\em JHEP} {\bf 05} (2006)  002,
\href{http://arxiv.org/abs/hep-ph/0602028}{{\tt arXiv:hep-ph/0602028}}.
%%CITATION = HEP-PH/0602028;%%.

\bibitem{Berger:2008cq}
C.~F. Berger, J.~S. Gainer, J.~L. Hewett, and T.~G. Rizzo, ``{Supersymmetry
  Without Prejudice},''
  \href{http://dx.doi.org/10.1088/1126-6708/2009/02/023}{{\em JHEP} {\bf 02}
  (2009)  023},
\href{http://arxiv.org/abs/0812.0980}{{\tt arXiv:0812.0980 [hep-ph]}}.
%%CITATION = 0812.0980;%%.

\bibitem{AbdusSalam:2009qd}
S.~S. AbdusSalam, B.~C. Allanach, F.~Quevedo, F.~Feroz, and M.~Hobson,
  ``{Fitting the Phenomenological MSSM},''
\href{http://arxiv.org/abs/0904.2548}{{\tt arXiv:0904.2548 [hep-ph]}}.
%%CITATION = 0904.2548;%%.

\bibitem{Baer:2005bu}
H.~Baer, A.~Mustafayev, S.~Profumo, A.~Belyaev, and X.~Tata, ``{Direct,
  indirect and collider detection of neutralino dark matter in SUSY models with
  non-universal Higgs masses},'' {\em JHEP} {\bf 07} (2005)  065,
\href{http://arxiv.org/abs/hep-ph/0504001}{{\tt arXiv:hep-ph/0504001}}.
%%CITATION = HEP-PH/0504001;%%.

\bibitem{Bottino:2001dj}
A.~Bottino, F.~Donato, N.~Fornengo, and S.~Scopel, ``{Size of the neutralino
  nucleon cross-section in the light of a new determination of the pion nucleon
  sigma term},'' \href{http://dx.doi.org/10.1016/S0927-6505(02)00107-X}{{\em
  Astropart. Phys.} {\bf 18} (2002)  205--211},
\href{http://arxiv.org/abs/hep-ph/0111229}{{\tt arXiv:hep-ph/0111229}}.
%%CITATION = HEP-PH/0111229;%%.

\bibitem{Belanger:2008sj}
G.~Belanger, F.~Boudjema, A.~Pukhov, and A.~Semenov, ``{Dark matter direct
  detection rate in a generic model with micrOMEGAs2.2},''
  \href{http://dx.doi.org/10.1016/j.cpc.2008.11.019}{{\em Comput. Phys.
  Commun.} {\bf 180} (2009)  747--767},
\href{http://arxiv.org/abs/0803.2360}{{\tt arXiv:0803.2360 [hep-ph]}}.
%%CITATION = 0803.2360;%%.

\bibitem{belanger_id}
G.~B\'elanger {\it et al.} in preparation.

\bibitem{Zhang:2008pka}
Z.~Zhang, ``{Muon g-2: a mini review},''
\href{http://arxiv.org/abs/0801.4905}{{\tt arXiv:0801.4905 [hep-ph]}}.
%%CITATION = 0801.4905;%%.

\bibitem{mcmc_book}
D.~Mackay, {\it Information theory, Inference, and learning Algorithms}, Cambridge University Press,2003.

\bibitem{Allanach:2001kg}
B.~C. Allanach, ``{SOFTSUSY: A C++ program for calculating supersymmetric
  spectra},'' \href{http://dx.doi.org/10.1016/S0010-4655(01)00460-X}{{\em
  Comput. Phys. Commun.} {\bf 143} (2002)  305--331},
\href{http://arxiv.org/abs/hep-ph/0104145}{{\tt arXiv:hep-ph/0104145}}.
%%CITATION = HEP-PH/0104145;%%.

\bibitem{Belanger:2004yn}
G.~Belanger, F.~Boudjema, A.~Pukhov, and A.~Semenov, ``{micrOMEGAs: Version
  1.3},'' \href{http://dx.doi.org/10.1016/j.cpc.2005.12.005}{{\em Comput. Phys.
  Commun.} {\bf 174} (2006)  577--604},
\href{http://arxiv.org/abs/hep-ph/0405253}{{\tt arXiv:hep-ph/0405253}}.
%%CITATION = HEP-PH/0405253;%%.

\bibitem{Belanger:2006is}
G.~Belanger, F.~Boudjema, A.~Pukhov, and A.~Semenov, ``{micrOMEGAs2.0: A
  program to calculate the relic density of dark matter in a generic model},''
  \href{http://dx.doi.org/10.1016/j.cpc.2006.11.008}{{\em Comput. Phys.
  Commun.} {\bf 176} (2007)  367--382},
\href{http://arxiv.org/abs/hep-ph/0607059}{{\tt arXiv:hep-ph/0607059}}.
%%CITATION = HEP-PH/0607059;%%.

\bibitem{Misiak:2006zs}
M.~Misiak {\em et al.}, ``{The first estimate of B(anti-B --> X/s gamma) at
  O(alpha(s)**2)},''
  \href{http://dx.doi.org/10.1103/PhysRevLett.98.022002}{{\em Phys. Rev. Lett.}
  {\bf 98} (2007)  022002},
\href{http://arxiv.org/abs/hep-ph/0609232}{{\tt arXiv:hep-ph/0609232}}.
%%CITATION = HEP-PH/0609232;%%.

\bibitem{Abazov:2007ww}
{\bf D0} Collaboration, V.~M. Abazov {\em et al.}, ``{Search for squarks and
  gluinos in events with jets and missing transverse energy using 2.1 $fb^{-1}$
  of $p \bar{p}$ collision data at $\sqrt{s}$ = 1.96- TeV},''
  \href{http://dx.doi.org/10.1016/j.physletb.2008.01.042}{{\em Phys. Lett.}
  {\bf B660} (2008)  449--457},
\href{http://arxiv.org/abs/0712.3805}{{\tt arXiv:0712.3805 [hep-ex]}}.
%%CITATION = 0712.3805;%%.

\bibitem{Davier:gmuon}
M.~Davier, ``{g-2},''
\href{http://arxiv.org/abs/Talk presented at Tau'08, Novosibirsk,Russia.}{{\tt
  Talk presented at Tau'08, Novosibirsk,Russia.}}
%%%%.

\bibitem{Dedes:2001fv}
A.~Dedes, H.~K. Dreiner, and U.~Nierste, ``{Correlation of B/s --> mu+ mu- and
  (g-2)(mu) in minimal supergravity},''
  \href{http://dx.doi.org/10.1103/PhysRevLett.87.251804}{{\em Phys. Rev. Lett.}
  {\bf 87} (2001)  251804},
\href{http://arxiv.org/abs/hep-ph/0108037}{{\tt arXiv:hep-ph/0108037}}.
%%CITATION = HEP-PH/0108037;%%.


\bibitem{moriond_talk}
J. Lewis, talk presented at Moriond electrowek 2009, LaThuile, Italy, March 2009.

\bibitem{Dedes:2002zx}
A.~Dedes, H.~K. Dreiner, U.~Nierste, and P.~Richardson, ``{Trilepton events and
  $B/s \rightarrow \mu^+ \mu^-$: No-lose for mSUGRA at the Tevatron?},''
\href{http://arxiv.org/abs/hep-ph/0207026}{{\tt arXiv:hep-ph/0207026}}.
%%CITATION = HEP-PH/0207026;%%.

\bibitem{Mercadante:2005vx}
P.~G. Mercadante, J.~K. Mizukoshi, and X.~Tata, ``{Using b-tagging to enhance
  the SUSY reach of the CERN Large Hadron Collider},''
  \href{http://dx.doi.org/10.1103/PhysRevD.72.035009}{{\em Phys. Rev.} {\bf
  D72} (2005)  035009},
\href{http://arxiv.org/abs/hep-ph/0506142}{{\tt arXiv:hep-ph/0506142}}.
%%CITATION = HEP-PH/0506142;%%.

\bibitem{Kadala:2008uy}
R.~H.~K. Kadala, P.~G. Mercadante, J.~K. Mizukoshi, and X.~Tata,
  ``{Heavy-flavour tagging and the supersymmetry reach of the CERN Large Hadron
  Collider},'' \href{http://dx.doi.org/10.1140/epjc/s10052-008-0672-9}{{\em
  Eur. Phys. J.} {\bf C56} (2008)  511--528},
\href{http://arxiv.org/abs/0803.0001}{{\tt arXiv:0803.0001 [hep-ph]}}.
%%CITATION = 0803.0001;%%.

\bibitem{DeSanctis:2007td}
U.~De~Sanctis, T.~Lari, S.~Montesano, and C.~Troncon, ``{Perspectives for the
  detection and measurement of Supersymmetry in the focus point region of
  mSUGRA models with the ATLAS detector at LHC},''
  \href{http://dx.doi.org/10.1140/epjc/s10052-007-0415-3}{{\em Eur. Phys. J.}
  {\bf C52} (2007)  743--758},
\href{http://arxiv.org/abs/0704.2515}{{\tt arXiv:0704.2515 [hep-ex]}}.
%%CITATION = 0704.2515;%%.

\bibitem{Profumo:2004at}
S.~Profumo and C.~E. Yaguna, ``{A statistical analysis of supersymmetric dark
  matter in the MSSM after WMAP},''
  \href{http://dx.doi.org/10.1103/PhysRevD.70.095004}{{\em Phys. Rev.} {\bf
  D70} (2004)  095004},
\href{http://arxiv.org/abs/hep-ph/0407036}{{\tt arXiv:hep-ph/0407036}}.
%%CITATION = HEP-PH/0407036;%%.

\bibitem{Allanach:2007qk}
B.~C. Allanach, K.~Cranmer, C.~G. Lester, and A.~M. Weber, ``{Natural Priors,
  CMSSM Fits and LHC Weather Forecasts},''
  \href{http://dx.doi.org/10.1088/1126-6708/2007/08/023}{{\em JHEP} {\bf 08}
  (2007)  023},
\href{http://arxiv.org/abs/0705.0487}{{\tt arXiv:0705.0487 [hep-ph]}}.
%%CITATION = 0705.0487;%%.

\bibitem{Trotta:2008bp}
R.~Trotta, F.~Feroz, M.~P. Hobson, L.~Roszkowski, and R.~Ruiz~de Austri, ``{The
  Impact of priors and observables on parameter inferences in the Constrained
  MSSM},'' \href{http://dx.doi.org/10.1088/1126-6708/2008/12/024}{{\em JHEP}
  {\bf 12} (2008)  024},
\href{http://arxiv.org/abs/0809.3792}{{\tt arXiv:0809.3792 [hep-ph]}}.
%%CITATION = 0809.3792;%%.

\end{thebibliography}
\end{document}